\documentclass[aps, reprint]{revtex4-2}
\usepackage{graphicx}
\usepackage{amsmath}
\usepackage{upgreek}
\usepackage[english]{babel}
\usepackage{braket}
\usepackage{bbold}
\usepackage{hyperref}
\hypersetup{colorlinks=true,linkcolor=blue, filecolor=blue, urlcolor=blue, citecolor=blue}
\usepackage{color}
\definecolor{darkgreen}{rgb}{0.01, 0.75, 0.24}

\definecolor{darkred}{rgb}{0.72, 0.11, 0.17}

\begin{document}
\title{Observing coherence in an incoherent paramagnetic nitrogen spin bath}
\author{R. M. Goldblatt}
\author{A. M. Martin}
\author{A. A. Wood}
\email{alexander.wood@unimelb.edu.au}
\affiliation{School of Physics, University of Melbourne, Parkville Victoria 3010, Australia}
\date{\today}

\begin{abstract}
The unpolarized spin environment surrounding a central spin qubit is typically considered as an incoherent source of dephasing, however, precise characterization and control of the spin bath can yield a resource for storing and sensing with quantum states. In this work, we use nitrogen-vacancy (NV) centers in diamond to measure the coherence of optically-dark paramagnetic nitrogen defects (P1 centers) and detect coherent interactions between the P1 centers and a local bath of $^{13}$C nuclear spins. The dipolar coupling between the P1 centers and $^{13}$C nuclear spins is identified by signature periodic collapses and revivals in the P1 spin coherence signal. We then demonstrate, using a range of dynamical decoupling protocols, that the probing NV centers and the P1 spins are coupled to independent ensembles of $^{13}$C nuclear spins. Our work illustrates how the optically-dark P1 spins, despite being unpolarized, can be used to extract information from their local environment and offers new insight into the interactions within a many-body system.

\end{abstract}

\maketitle
\section{Introduction}
The study of how a central spin qubit interacts with a bath of surrounding spins features prominently in quantum sensing, where decoupling the sensing qubit from the noisy environment is the principal concern~\cite{de_lange_universal_2010, de_lange_controlling_2012, bar-gill_suppression_2012, tyryshkin_electron_2012, bauch_ultralong_2018}. In quantum information processing (QIP), precise knowledge and control of the spin environment facilitates a resource for storing, entangling and measuring quantum states~\cite{cai_large-scale_2013, unden_coherent_2018, hermans_qubit_2022, cujia_parallel_2022}. Diamond is one of the most widely studied wide-bandgap semiconductors hosting spin qubits with applications in quantum sensing and information processing, and the optically-active nitrogen-vacancy (NV) center \cite{doherty_nitrogen-vacancy_2013} can be used to measure and control the optically-inactive, many-body spin environment surrounding it. For natural abundance diamond, the dominant spin environment is composed of spin-1/2 $^{13}$C nuclei and substitutional neutrally-charged nitrogen impurities (P1 centres) and much effort has been invested in schemes to eliminate the influence of these spins on the NV coherence for quantum sensing~\cite{barry_sensitivity_2020}. The $^{13}$C spins around individual NV centers, however, weakly couple to magnetic fields, and as such possess intrinsically long spin coherence times that are highly amenable to schemes for storing, processing and retrieving quantum information~\cite{dutt_quantum_2007, neumann_multipartite_2008, maurer_room-temperature_2012, taminiau_universal_2014, abobeih_atomic-scale_2019, abobeih_fault-tolerant_2022, hermans_qubit_2022, cujia_parallel_2022, van_de_stolpe_mapping_2023}. 

Similarly, the ability to control, polarize and entangle P1 centers near NVs has positioned the P1 itself as a potential platform for quantum information processing  \cite{laraoui_approach_2013, belthangady_dressed-state_2013, degen_entanglement_2021}. Additionally, the tunable coupling of the P1 center to both control fields~\cite{goldblatt_tunable_2022} and other spins in the diamond can be utilized for enhanced sensing schemes, in which the P1 electron spins act as useful ancillae or reporter sensors \cite{goldstein_environment-assisted_2011, knowles_demonstration_2016, cooper_environment-assisted_2019}. While independent understanding of the NV-$^{13}$C interaction and NV-P1 interaction has been the focus of much work, the full NV-$^{13}$C-P1 interaction, in particular the coupling between the P1 spins and their own $^{13}$C spin bath has been somewhat overlooked, despite several works revealing nontrivial effects, such as P1-$^{13}$C mediated hyperpolarisation and spin torques~\cite{wunderlich_optically_2017, pagliero_multispin-assisted_2018, zangara_mechanical_2019, henshaw_carbon-13_2019}.

In this work, we use an ensemble of NV centers in diamond to probe the coherence of the optically-dark P1 centers, and detect coherent interactions between the P1 spins and $^{13}$C nuclei, despite having no spin polarization of either. We observe periodic collapses and revivals in the coherence of the P1 spins, and show that these modulations are induced by coherent interactions between the P1 centers and a local bath of $^{13}$C nuclear spins \cite{childress_coherent_2006}. Interestingly, the local $^{13}$C spins surrounding the probing NV centres are independent from the P1-coupled nuclear spin bath, which we show by varying the degree to which dynamical decoupling protocols isolate the NV centre from the combined spin bath. Our observations are well explained by numerical simulations of the P1-$^{13}$C spin bath. These results reveal a surprisingly coherent interaction hidden within an unpolarized many-body spin bath, and will be of interest to schemes that employ P1 spins as reporter spins~\cite{zhang_reporter-spin-assisted_2023} or as probes of many body physics~\cite{davis_probing_2023} or to serve as a remote electronic spin node in an NV-$^{13}$C quantum information processing network~\cite{van_de_stolpe_mapping_2023}. Our results should also further stimulate the renewed theoretical attention devoted to the spin physics of P1 centers recently~\cite{park_decoherence_2022, onizhuk_bath-limited_2023}.

\begin{figure*}[tbp]	\center{\includegraphics[width=\textwidth,keepaspectratio]{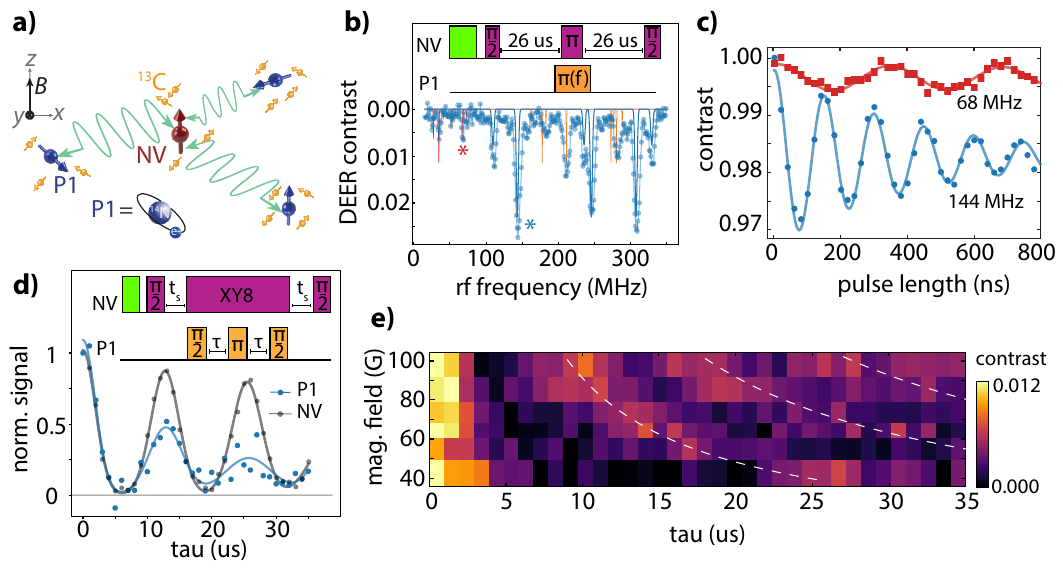}} 
	\caption{\label{fig1} Measuring the dynamics of the P1 spin bath using an ensemble of NV centers. a) Schematic of the system: an ensemble of NV centers interacts with an ensemble of P1 centers. Both the NV centers and P1 centers are embedded within a bath of $^{13}$C nuclear spins. An external magnetic field $B$ is aligned with the NV symmetry axis along the [111] crystallographic direction. b) Pulse sequence of DEER measurement (upper) and the spectrum at $B = 72$G. The calculated resonant frequencies for particular transitions are displayed under the experimental signal, color-coded by transition type: blue for electron spin transitions, orange for double quantum transitions, and red for nuclear spin transitions. (c) Coherent Rabi oscillations of the P1 bath spins using rf pulses of 144\,MHz (blue), which drives an electron transition, and 68\,MHz (red), which drives a hybridized electron-nuclear spin transition. (d) NV spin-echo signal at $B = 72$\,G (gray) shows the characteristic periodic collapses and revivals of coherence associated with an NV center coupled to a precessing $^{13}$C spin bath. Spin-echo measurement on the P1 spin bath at 72\,G (blue) where an XY8 dynamical decoupling sequence is applied to the NV ensemble during the coherence measurement (inset). Solid lines denote Gaussian fits to the data (points). (e) Color map of the P1 electron spin echo signals for increasing magnetic field strength. The dashed lines show the integer multiples of the $^{13}$C Larmor precession period for each magnetic field, which are given by $\tau = \tau_{L} = (\gamma_{\textrm{13C}} B)^{-1}$.}
\end{figure*}

\begin{figure}[tbp]	\center{\includegraphics[width=0.5\textwidth,keepaspectratio]{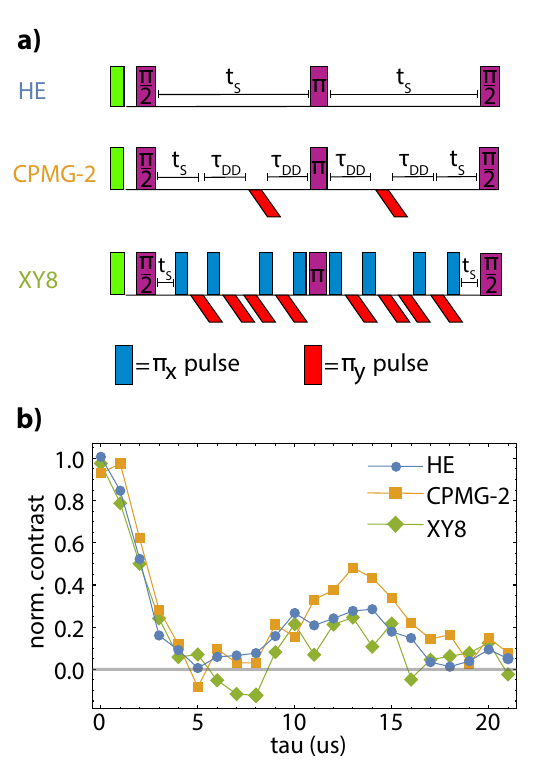}} 
	\caption{\label{fig2} P1 spin-echo measurements under different dynamical decoupling schemes applied to the NV. a) Pulse sequences used in the experiments to decouple the NV from its environment, including Hahn-echo (top), CPMG-2 (middle) and XY8 (bottom). The blue pulses denote rotations about the $x$-axis and the red pulses are orthogonal, representing rotations about the $y$-axis. In all sequences, the time $t_s$ is the period in which the NV is sensitive to the dipolar field from the P1 spin bath. b) Measurements of the P1 electron spin-echo signal using each of the NV dynamical decoupling sequences at a magnetic field of 72G. All three signals show a revival at the Larmor period originating from $^{13}$C nuclear spins. The variation in the signal-to-noise ratio between the signals is attributed to the change in the duty cycle for each dynamical decoupling sequence and the accumulation of pulse fidelity errors.}
\end{figure}

\section{Experiment}
We consider an ensemble of \mbox{spin-1} NV centers interacting with an ensemble of P1 centers and a bath of $^{13}$C nuclear spins, represented by the schematic in Figure \ref{fig1}(a). We use a $\langle111\rangle$-cut diamond sample with a 1.1\% natural abundance of $^{13}$C and a nitrogen concentration of 1 ppm as reported by the manufacturer (Delaware Diamond Knives). Our experimental setup is described in greater detail in Ref. \cite{goldblatt_tunable_2022}. A magnetic field is applied along the [111] crystallographic axis, which we denote as the $\hat{z}$-axis, spectrally selecting out a single NV orientation class. The magnetic field lifts the degeneracy of the $\ket{m_S = \pm1}$ spin states of the NVs, allowing us to work in the effective spin-$\frac{1}{2}$ subspace $\{\ket{m_S = 0}, \ket{m_S = -1}\}$. Microwaves (mw) for NV driving and radiofrequency (rf) for P1 control are synthesized using an arbitrary waveform generator (AWG, Tabor Proteus P2584M). Microwave waveforms are generated via single-sideband IQ modulation of a local oscillator ($f_\text{IQ} = 50\,$MHz), while rf waveforms are directly generated on the AWG. The signals are then amplified by two separate amplifiers and applied via wires with a diameter of 20\,$\upmu$m, arranged in a cross and in contact with the diamond surface. For the magnetic field strengths used in this work, the NVs and P1s are highly detuned from each other, which enables independent control of the spins and ensures strong suppression of spin-exchange interactions between the spin species present near the ground and excited state level anticrossings~\cite{wunderlich_optically_2017, pagliero_multispin-assisted_2018}. 

Three substitutional nitrogen charge states are known~\cite{ashfold_nitrogen_2020}, but only the neutral charge state (N$^{0}$) exhibits magnetic resonance activity and is referred to as the P1 centre \cite{deak_formation_2014, ulbricht_single_2011}, consisting of an electron spin ($S = 1/2$) associated with an unpaired electron coupled to a $^{14}$N nuclear spin ($I = 1$). The P1 center exhibits a static Jahn-Teller (JT) distortion \cite{davies_dynamic_1979, davies_jahn-teller_1981, ammerlaan_reorientation_1981}, which results in four possible symmetry axes due to an elongation of one of the N-C bonds. At room temperature, a given P1 centre will randomly reorient between all possible JT axes at a rate of $\sim$0.3\,kHz \cite{ammerlaan_reorientation_1981, doherty_towards_2016}. In our experiments, we probe a large ensemble of P1s, and as a result observe a temporal average of all JT orientation classes. P1 centers oriented along the same crystallographic axis as the NV orientation class that we target with microwaves (i.e. along the $\hat{z}$ axis) are parallel to the magnetic field and will be referred to as `on-axis', and the other three degenerate P1 orientations make an angle of $\theta = 109.5^{\circ}$ to the $\hat{z}$-axis and will be referred to as `off-axis'. In this work, we use the off-axis P1 centers due to the three times larger signal to noise, though everything we discuss herein is equally applicable to the on-axis classes. 

For a single P1 center with the JT axis along $z'$, the interaction Hamiltonian is given by:
\begin{align}
\begin{split} \label{HP1}
H_{P1} ={}& \gamma_e \vec{\boldsymbol{B}} \cdot \vec{\boldsymbol{S}} \, \, \gamma_N \vec{\boldsymbol{B}} \cdot \vec{\boldsymbol{I}} + \, A_{\parallel} S_{z'} I_{z'} \\
& + A_{\perp} (S_{x'} I_{x'} + S_{y'} I_{y'}) \, + \, Q I_{z'}^2 \, ,
\end{split}
\end{align}
where $\gamma_e / 2\pi = -2.8$\,MHz/G and  $\gamma_N / 2\pi = 307.7$\,Hz/G are the gyromagnetic ratios of the electron and the $^{14}$N nuclear spin. The hyperfine interactions between the electron ($S$) and nuclear spin ($I$) is defined in terms of an axial coupling term, $A_{\parallel} / 2\pi = 114$\,MHz, and a transverse component, $A_{\perp} / 2\pi = 81.34$\,MHz \cite{degen_entanglement_2021}. The zero-field splitting between the nuclear spin states is quantified by the nuclear quadrupole coupling term, $Q/ 2\pi = -4.2$\,MHz. 

We use double electron-electron resonance (DEER) spectroscopy to characterize the spin bath surrounding the NV centers. The DEER sequence, depicted in Figure \ref{fig1}(b) consists of a spin-echo pulse sequence applied to the NV and an rf $\pi$-pulse, which selectively recouples resonant P1 spins. An exemplar spin-echo signal for the NV ensemble at a magnetic field strength of 72\,G is shown in Figure \ref{fig1}(d). The characteristic collapses and revivals of the NV spin-echo signal indicate that the $^{13}$C nuclear spin bath is the dominant environment of the NV ensemble \cite{childress_coherent_2006}, a typical situation for electronic and low-nitrogen ($\sim$ppm) standard grade diamond samples~\cite{bauch_decoherence_2020}. The signature of P1 effects on the NV coherence are apparent in the damping of the revival amplitudes, which results in $T_2 = 150\,\upmu$s, even when the magnetic field is parallel to the NV axis, just discernible in this sample over the timescales we study yet much faster than homonuclear flip-flop interactions between $^{13}$C spins~\cite{maze_electron_2008, stanwix_coherence_2010}. On a practical level, the interactions between the NVs and the $^{13}$C spin bath also restrict the spin-echo visibility to the $^{13}$C-induced revivals. Accordingly, the free evolution time of the DEER sequence is fixed at a $^{13}$C-induced revival time in order to ensure signal visibility.

Figure \ref{fig1}(b) shows the DEER signal at $B = 72$\,G as a function of rf frequency, where the contrast is defined as the difference between the NV spin-echo signal with and without the rf pulse. The expected frequencies for all possible P1 spin transitions, calculated from the eigenvalues of Eq. \ref{HP1} are also shown. We observe the six main P1 electron-spin transitions for the on-axis and off-axis orientations. The larger amplitudes of the off-axis P1s arises from the three degenerate orientations. We also observe several additional spectral features, corresponding to transitions between hybridized electron-nuclear spin states \cite{goldblatt_tunable_2022}. 

Varying the length of the resonant rf pulse traces out Rabi oscillations of the P1 spins, shown for one of the off-axis electron transitions (144\,MHz, $\ket{m_{S} = 1/2, m_{I} = -1} \leftrightarrow \ket{-1/2, -1}$) and one of the hybridized electron-nuclear transitions (68\,MHz, $\ket{-1/2, -1} \leftrightarrow \ket{-1/2, 0}$) in Figure \ref{fig1}(c). The decreased amplitude and Rabi frequency of the oscillation for the hybridized transition is well explained by the augmentation of the effective gyromagnetic ratio induced by the mixing of the electron and nuclear spin states \cite{goldblatt_tunable_2022}. 

\section{P1 spin-echo coherence measurement}

To probe the coherence of the P1 spins, we alter the DEER sequence so that the single rf pulse is replaced with a $\pi/2$-$\tau$-$\pi$-$\tau$-$\pi/2$ spin-echo pulse sequence applied to the P1 spins and the spin-echo sequence applied to the NV is replaced with a dynamical-decoupling (DD) sequence. Dynamical decoupling schemes protect the coherence of a central spin from environmental noise by repeatedly flipping the spin state such that the effect of the noise is averaged out \cite{viola_dynamical_1999, de_lange_universal_2010, ryan_robust_2010, naydenov_dynamical_2011}. In our case, we use a dynamical decoupling protocol on the NV ensemble to effectively decouple the NV from any time-dependent P1 dynamics occurring over the extended DEER sequence that now coherently manipulates the P1 spins. We replace the single $\pi$-pulse from the spin-echo sequence previously used on the NVs with a chain of dynamical decoupling pulses. We also ensure that the net refocusing $\pi$-rotation of the NV spins is preserved so that the P1 state following spin manipulation can still be measured. The measurement sequence can be understood by considering the two periods of free evolution for the NVs, $t_s$, on either side of the DD sequence as sensing periods, in which the NVs accumulate a phase dependent on the P1 spin bath, before and after the P1 spin echo sequence. During the dynamical decoupling time, the NV is prevented from coupling to the P1 spins by rapidly flipping its spin state, allowing the P1 dynamics to be probed in isolation. The amplitude of the NV spin-echo signal will depend on the difference between the dipolar field produced by the P1s before and after the rf pulse sequence, and therefore, will serve as a measure of the coherence of the spin bath. To ensure that the P1 spin echo dynamics are measured unambiguously, after readout of the NV fluorescence we perform an identical measurement sequence with the final $\pi/2$ pulse of the \emph{P1 spin-echo sequence} set to $-180^\circ$ relative to the first measurement, and take as our overall signal the difference between these two signals. In this manner, only transverse spin dynamics of the P1, which depend on the phase of the final rf pulse, contribute to the signal.

A spin-echo measurement of the P1 electron spins as a function of the free evolution time $\tau$ is shown in Figure \ref{fig1}(d), where an XY-8 dynamical decoupling sequence has been applied to the NV ensemble \cite{gullion_new_1990}. We observe a periodic modulation in the spin-echo signal, which is consistent with the collapses and revivals observed in a typical spin-echo measurement for an NV center in the presence of precessing $^{13}$C spin bath [Fig \ref{fig1}(d)]. To verify that the modulation in the P1 spin-echo signal is indeed induced by the precessing $^{13}$C nuclear spin bath, we repeat the coherence measurement across a range of external magnetic fields strengths [Fig. \ref{fig1}(e)]. For each magnetic field, we adjust the sensing period, $t_s$, of the NV sequence to match a $^{13}$C-induced revival. In Figure \ref{fig1}(e) we observe that across a range of magnetic field strengths, the revivals in the P1 spin echo signal precisely match integer multiples of the $^{13}$C Larmor precession period. Exponential fits to the signals in Fig. \ref{fig1}(e) yield P1 coherence times between $T_2 = 30-40\,\upmu$s.

\begin{figure}	\center{\includegraphics[width=0.5\textwidth,keepaspectratio]{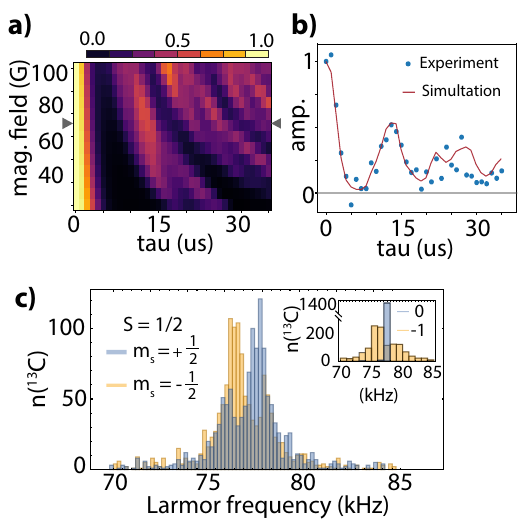}} 
	\caption{\label{fig3} Numerical simulations of the P1 spin-echo signals in the presence of a $^{13}$C nuclear spin bath. a) Simulated spin-echo signals for the P1 electron spin as a function of magnetic field strength. b) The simulated spin-echo signal at $B = 72$\,G (indicated by the gray arrows in (a)) compared to the experimental signal measured at the same magnetic field. c) Distribution of nuclear spin precession frequencies when coupled to a central P1 electron spin in the $m_s = +1/2$ state (blue) and $m_s = -1/2$ state (yellow). The distribution of the Larmor frequencies for the same bath of $^{13}$C spins when coupled to the spin-1 NV electron spin (inset).}
\end{figure}

\section{Variable decoupling of the NV}

We previously invoked dynamical decoupling of the NV in order to disentangle any possible effect from the P1 coherence evolution. Next, we measured the coherence signal of the P1s while applying different dynamical decoupling protocols to the NV ensemble in order to examine the effect of the NV evolution on the P1-$^{13}$C dynamics. Figure \ref{fig2}(a) depicts three of the sequences used in order of the total number of pulses: Hahn-echo, CPMG-2 \cite{carr_effects_1954, meiboom_modified_1958} and the XY8 sequence, outlined previously. Each sequence is applied symmetrically about the central $\pi$-pulse. The P1 spin-echo signal for each DD sequence is shown in Figure \ref{fig2}(b), at a magnetic field of $B = 72$\,G. In each case, following the initial collapse in the signal, we observe a revival centered at the Larmor period, $\tau_L = 12.96\,\upmu$s. The presence of the revival in the P1 spin echo signal is independent of dynamical decoupling schemes, which further verifies that the signal directly results from the interactions between the P1 spin bath and the precessing $^{13}$C nuclear spins. The signal-to-noise ratio of the measurements is the only apparent difference between the DD schemes, which can be easily understood in terms of the different duty cycles of the experiments and compounding pulse fidelity errors. Notably, the revivals in the P1 spin echo persist in the absence of dynamical decoupling on the NV, which is evident in the Hahn-echo signal shown in Fig. \ref{fig2}(b). The free evolution of the NVs, therefore, does not significantly perturb the coherent interactions between the P1 centers and their local $^{13}$C nuclear spin bath, which induce the revivals in the spin-echo signal.

To understand why it is not necessary to decouple the NV entirely, we consider the average NV-P1 separation. The average separation between an NV and its $k$th neighbour P1 center is given by: 
\begin{equation}
\langle r_k \rangle = \biggl( \frac{4 \pi n}{3} \biggr)^{- \frac{1}{3}} \, \frac{\Gamma(k + \frac{1}{3})}{\Gamma(k)}
\end{equation}
where $n$ is the density of P1 centers and $\Gamma$ is the gamma function, such that $\Gamma(k) = (k-1)!$ \cite{hall_analytic_2014, hall_detection_2016}. For our sample, we calculate an average separation distance between an NV and its nearest P1 of $\langle r_{1} \rangle = 16.9 \pm 0.6$\,nm, based on the concentration range of N$^{0}$ in the sample (See Appendix A). Approximating the NV and P1 electron spins as point dipoles, the average dipole-dipole interaction between the spins is given by
\begin{equation}
\langle V_{\textrm{dip}} \rangle = \frac{\upmu_0 \gamma_{e} \gamma_{e} (1 - 3\cos^2\langle\theta\rangle)}{4 \pi \langle r \rangle ^3} ,
\end{equation}
where $\upmu_0$ is the vacuum permittivity, $\langle r \rangle$ is the average separation between the dipoles and $\theta$ is the angle between the two spins. From this equation, we calculate an average coupling strength between the NV and its nearest P1 of $\langle V_{dip} \rangle = 5.4 \pm 0.6$\,kHz, which corresponds to a coupling timescale of $185 \pm 20\,\upmu$s, which is significantly longer than the P1 measurement time though commensurate with the $T_{2}$ time of the diamond. This implies that the NV centers and P1s interact with distinct $^{13}$C nuclear spin baths. The `frozen core radius' for the NV-$^{13}$C system, a term denoting the region in which the nuclear spin dynamics are dominated by their hyperfine interactions with the electron spin, is estimated to be between $2.7 - 3.8 \textrm{nm}$, which supports our finding that the NVs and P1s in our sample interact with distinct $^{13}$C spin baths \cite{onizhuk_bath-limited_2023}.

\begin{figure}[tbp]	\center{\includegraphics[width=0.49\textwidth,keepaspectratio]{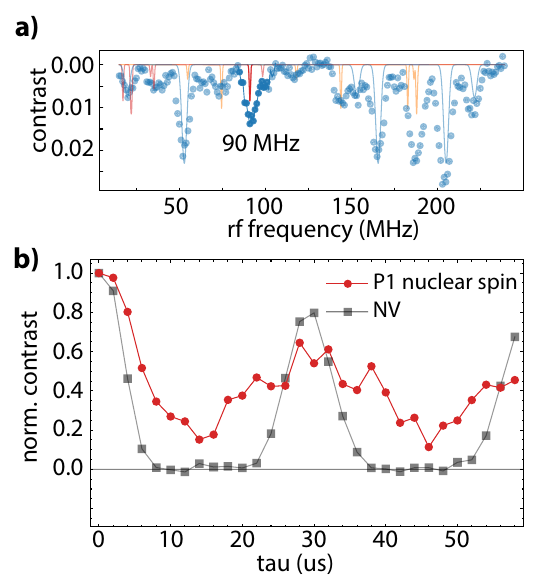}} 
	\caption{\label{fig4} Hybridized electron-nuclear spins. a) DEER spectrum at $B = 32$G, where we target the highlighted nuclear transition (red line) at 90\,MHz. b) The spin-echo signal for the P1 nuclear spin (red) compared to the standard NV spin-echo signal (gray). The agreement between the two signals shows that at low magnetic fields hybridized P1 electron-nuclear spins can also detect $^{13}$C dynamics.}
\end{figure}

The isolation of the P1s from the NV centers enables us to construct a concise and simplified model to simulate the system. Since the NVs are effectively acting as weakly-perturbative observers of the P1 spin states, we are able to reproduce the experimental results by directly simulating the P1 dynamics without explicit inclusion of the NV measurement process in the presence of a $^{13}$C spin bath. We compared our experimental results to numerical simulations of the P1 spin-echo signal generated by adapting the disjoint-cluster methods of Ref. \cite{maze_electron_2008, wood_anisotropic_2021}, which were developed for the NV-$^{13}$C case. A bath of $^{13}$Cs is randomly distributed around a central P1 spin and the $^{13}$Cs are grouped such that there are strong interactions between spins within a group and negligible coupling between groups. The maximum group size is set at $g = 3$ in order to include spin correlations up to third order. The simulated signal is computed as the coherence of the system and the total spin-echo signal for a given bath of $^{13}$Cs is $S_T = \prod_{G} S_{G}$, where $S_G$ is the normalized spin-echo signal of the P1 for a particular group $G$ of nuclear spins. To match the experimental conditions in which we use an ensemble of NVs to address multiple P1 centers, we simulate ensemble averaging by computing the average signal, $S = 1/N_{\textrm{B}} \sum_{N_{\textrm{B}}} S_T$, for $N_{\textrm{B}}$ distributions of $^{13}$C spin baths. Each bath contains $125$ $^{13}$C nuclear spins and the simulation results are averaged over $N_{\textrm{B}} = 20$ spin baths. 

The calculation results are shown in Figure \ref{fig3}(a) and closely reproduce our experimental results. A line out at 72\,G (indicated by the gray arrows) is shown in Fig. \ref{fig3}(b) alongside the experimental signal at this magnetic field to highlight the fidelity of the simulated result. The concordance between the simulations and the experimental results, which infer the state of the P1s from the NV signal, confirms that the NVs are reliable sensors of the evolution of the P1s. In both experiment and simulation, the damping of the P1 spin-echo signal is significantly faster than the NV signal in the presence of the $^{13}$C nuclear spin bath. Additionally, we observe broader revivals in the P1 signal, which in some instances feature split peaks as in the second revival in Fig. \ref{fig3}(b). These features can also be seen in the density plot in Fig. \ref{fig3}(a). 

The differences between the NV and P1 spin-echo signals can be attributed to the differences in the coupling of the spins to the precessing nuclear spin bath. The Larmor precession frequency of the $i$th nuclear spin depends on the spin state of the proximal electronic spin $S$, i.e.
\begin{equation}
\omega_{L}^{i} = |E^{i}_{m_{S}, \uparrow} - E^{i}_{m_{S}, \downarrow}|, 
\end{equation}
where $m_S = 0, \pm1$ for the NV and $\pm 1/2$ for the P1 (See Appendix B). We can then compute the distribution of Larmor frequencies of the $^{13}$C spins for an NV central spin or a P1 central spin. Figure \ref{fig3}(c) shows the distribution of  Larmor precession frequencies in a $^{13}$C spin bath dependent on the P1 electron spin state and the NV electron spin state (inset). Unlike the NV, which during a spin-echo experiment is in a superposition of magnetic ($\ket{\pm 1}$) and non-magnetic ($\ket{0}$) states, the P1 electron spin does not possess a non-magnetic state. The free precession of the $^{13}$C nuclear spins, therefore, is always influenced by the P1 spin state, which leads to the faster damping in revival amplitude. The broadening of the revivals in the P1 signal can be attributed to the spread of $^{13}$C precession frequencies when coupled to both the P1 spin states [Fig. \ref{fig3}(c)]. Additionally, the splittings in the peaks of the revivals at increasing $\tau$ times may be explained by the separation between the peaks in the frequency distribution for the $\ket{m_s = +1/2}$ and $\ket{m_s = -1/2}$ P1 states. 

\section{Spin-echo of P1 nuclear spin states}

We also examined the spin-echo signal of the P1 nuclear spins using our dynamical decoupling techniques. At sufficiently low magnetic fields, where $\gamma_{E} B \sim A_{\perp}$, the mixing between the P1 electron and nuclear spins enables detection and coherent control of the nuclear spin state \cite{goldblatt_tunable_2022, degen_entanglement_2021}. Figure \ref{fig4}(a) shows the DEER spectrum at a magnetic field of $B = 32$\,G, where we highlight the resonance at 90\,MHz that corresponds to the nuclear spin transition $\ket{m_S = +1/2, m_I = -1} \leftrightarrow \ket{+1/2, 0}$. The spin-echo signal for the hybridized electron-nuclear spin transitions shows a revival at the $^{13}$C Larmor period and is consistent with the modulation in the NV spin-echo signal at the same magnetic field strength [Fig. \ref{fig4}(b)]. The observed spin-echo signal bares many similarities to what we have presented earlier for the purely-electronic transitions of the P1, with subtle differences: for the hybridized-nuclear spin transitions, the gyromagnetic ratio is set by the magnetic field strength, and while many times greater than the bare $^{14}$N gyromagnetic ratio $\gamma_N$ is less than that of the predominantly electron-like transition ($\gamma_{N'} = 158$\,kHz). Thus, the P1 spin echo damps slower, with a longer $T_2$ as a result ($70 \pm 3\,\upmu$s). The second feature, which is also a direct consequence of the reduced gyromagnetic ratio is the lack of sharp, split-revivals prevalent in electronic transition experimental data and simulations.

\section{Discussion and outlook}
The presence of P1 spins in diamond is essentially unavoidable. Materials processing steps can eliminate to a large degree the effects of $^{13}$C spins on NV center qubits, but the necessity of nitrogen incorporation during growth or implantation for NV generation -- and the low N to NV conversion ratios -- makes the presence of P1 centers ubiquitous. While many schemes have been developed to suppress the deleterious effects of P1 centers on NV spins, an alternative may be to employ the local environment of P1 spins as reporters~\cite{zhang_reporter-spin-assisted_2023}, which typically outnumber the NVs by an order of magnitude or more and occupy a larger sensing volume. A full understanding of how P1s behave is thus a necessary prerequisite in advancing them as a potential platform for quantum sensing. Our observations of $^{13}$C revivals in the P1 spin-echo signal reveals coherent interactions within the disordered, unpolarized spin bath. We demonstrate the P1 centers can be utilized as quantum sensors of their local environment, specifically, for the detection of precessing nuclear spins, one of the earliest such landmark demonstrations of the NV center~\cite{childress_coherent_2006}. Extensions to more sophisticated pulse sequences, such as correlation spectroscopy~\cite{laraoui_high-resolution_2013} capable of probing phase transitions in the $^{13}$C-P1 spin system, can be readily implemented.

Further study could focus on samples with lower or higher densities of N, in particular the threshold between P1- and $^{13}$C-dominated coherence for either species~\cite{bauch_decoherence_2020}, to examine the effects of overlapping the $^{13}$C spin baths coupled to an NV and P1 center. Another future avenue for measurements could be where the angle the magnetic field makes to the P1 axis is varied. Here, the coherence time of the NV center was observed to drop significantly when the magnetic field was misaligned from the NV axis~\cite{stanwix_coherence_2010}, inducing spin mixing of the NV and consequently anisotropic variation of the hyperfine coupling with the nearby $^{13}$C spins. For the P1, such effects will be reduced as the electron and nuclear spins will align more closely to the magnetic field direction. Indeed, the typical $T_{2,\text{P1}}\sim 10-30\,\upmu$s we observe in this work is for P1 spins with JT axes 109\,$^\circ$ to the applied magnetic field. Such a system may be amenable to magic-angle spinning schemes~\cite{hubrich_magic-angle_1997} to decouple the electron and nuclear interactions, which are difficult to realize with the NV due to the large zero-field splitting~\cite{wood_anisotropic_2021}. 

The presence of $^{13}$C revivals also confers sensitivity to physical rotation of the diamond, as the nuclear Larmor precession frequency changes based on the speed and direction of rotation~\cite{wood_magnetic_2017}. When combined with the additional spin of the P1 coupling, such sensitivity could be used to probe frame- and spin-dependent inertial forces in rapidly rotating diamonds~\cite{jin_quantum_2023}. 

Coherent interactions between the P1 and nearby $^{13}$C spins may also provide the basis for expanded spin-based quantum networks. Quantum registers comprised of nuclear spins coupled to a central electron spin, such as the NV-$^{13}$C system in diamond, are prospective platforms for quantum storage and quantum information processing given the weak coupling of the nuclear spins to their environment. The size of these nuclear spin networks, however, is limited by the magnetic dipolar coupling between the nuclear spins and the central electron spin. Recent experiments have proposed utilizing networks of coupled nuclear spins \cite{van_de_stolpe_mapping_2023} and chains of electron spin defects \cite{ungar_identification_2023} to engineer larger quantum registers. The results in this work could be utilized to combine these approaches to construct chains of NV and P1 electron spins that can in turn access local independent baths of $^{13}$C nuclear spins. This would enable distant $^{13}$C spins to be connected to the optically-active NVs through the P1 electron spin, in order to expand the volume of current NV-$^{13}$C based quantum registers.

\section*{Acknowledgements}
\vspace{-0.3cm} This work was supported by the Australian Research Council Discovery Scheme (DP190100949, DE210101093). R. M. G. was supported by a Research Training Program Scholarship.

\section*{Appendix A}
\begin{figure}	\center{\includegraphics[width=0.5\textwidth,keepaspectratio]{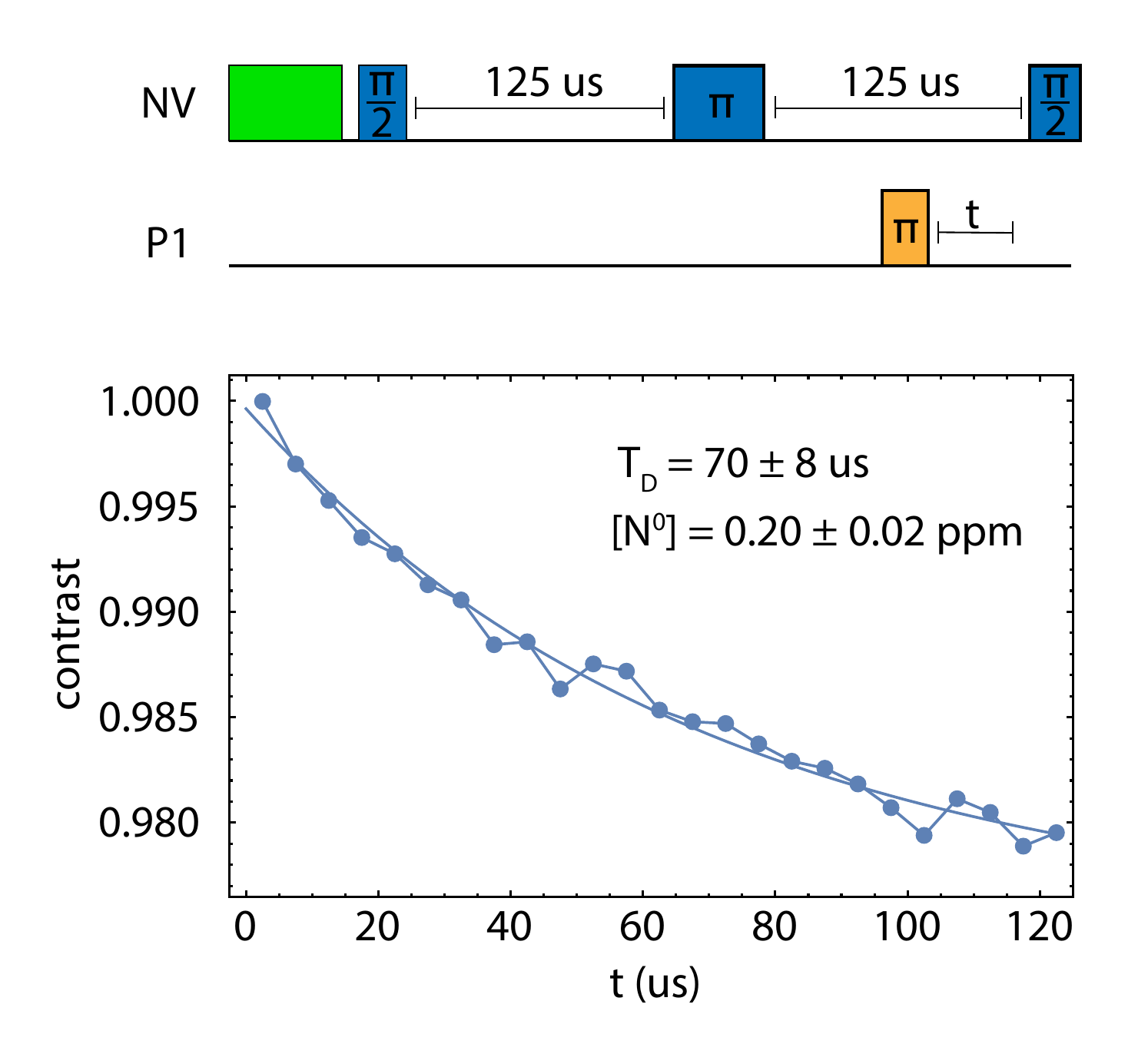}} 
	\caption{\label{p1conc} Determining the P1 concentration of the diamond sample. DEER signal as a function of the delay time $t$ using 70\,ns rf pulses of 144\,MHz, for $\tau_{SE} = 125\,\upmu$s. The solid line indicates an exponential decay fit with a time constant $T_D = 70 \pm 8 \upmu$s, which corresponds to a P1 density [N$^{0}$] = $0.2 \pm 0.02$\,ppm.}
\end{figure}

\vspace{-0.3cm}  \textbf{P1 Concentration.} The approximate concentration of nitrogen in the diamond sample can be determined using the DEER sequence shown in Figure \ref{p1conc}, where the timing of the P1 $\pi$-pulse relative to the final $\pi/2$ pulse is varied. When the time between the pulses is small, i.e. as $t \rightarrow 0$, the NV ensemble is insensitive to the change in the magnetic field induced by the resonant $\pi$-pulse and the NV spin-echo signal is at a maximum. As the time, $t$, increases, the measured signal will vary according to the coupling between the NV spin and the P1 spins in its environment. The measured signal is shown in Fig. \ref{p1conc}, where the single-exponential response indicates that each NV center is coupled to multiple P1 centers \cite{grotz_sensing_2011}. We extract a time constant of $T_D = 70 \pm 8 \upmu$s from an exponential decay fit and use this value to derive an approximate P1 concentration of [N$^0] = 0.2 \pm 0.02$ ppm \cite{salikhov_theory_1981, stepanov_determination_2016}. In similar diamond samples, the concentration of N$^{+}$ is typically 1-10 times the concentration of N$^{0}$ \cite{edmonds_production_2012}. Therefore, the P1 concentration that we have calculated is consistent with the manufacturer's specification of a total nitrogen concentration of [N] = 1 ppm.

\section*{Appendix B} \label{appB}
\vspace{-0.3cm} \textbf{Larmor Frequency Calculations.} In the presence of an external magnetic field, $B$, the $^{13}$C spins precess at a frequency given by $\omega_L = \gamma_{\textrm{13C}} B$, with $\gamma_{\textrm{13C}}/2\pi = 1071.5$\.Hz/G. The precession of a $^{13}$C spin in diamond will be influenced by its coupling to proximal electron spins, as the electron spin produces a large local magnetic field. Therefore, the $^{13}$C precession frequency will be conditional on the state of the electron spin and its relative position. The differences in the NV and P1 spin-echo signals, outlined in the main text, can be explained by the differences in the coupling of the spins to the $^{13}$C spin bath. In the main text, we calculate the distribution of Larmor frequencies for a bath of $^{13}$C spins surrounding an NV center and a P1 center using the eigenenergies of the Hamiltonians describing those respective systems.

The Hamiltonian describing the interaction between an NV (assumed to lie along the $z$-axis) and a single $^{13}$C is:
\begin{align} 
\begin{split} \label{NV-C13}
H_{\textrm{NV-13C}} ={}& D S_{z}^2 + \gamma_e \vec{\boldsymbol{S}} \cdot \vec{\boldsymbol{B}} \\
& + \gamma_{\textrm{13C}} \vec{\boldsymbol{I_C}} \cdot \vec{\boldsymbol{B}} + \vec{\boldsymbol{S}} \cdot \boldsymbol{A} \cdot \vec{\boldsymbol{I_C}},
\end{split}
\end{align}
where $D = 2.87$\,GHz is the zero-field splitting of the NV electron spin ($S$), $\gamma_e/2\pi=-2.8$\,MHz/G is the electron gyromagnetic ratio and $\boldsymbol{A}$ is the hyperfine tensor describing the interaction between the NV and the $^{13}$C nuclear spin ($I_C$). 

The Hamiltonian describing the P1-$^{13}$C system is:
\begin{equation}
\label{P1-C13}
H_{\textrm{P1-13C}} =  H_{P1} +  \gamma_{\textrm{13C}} \vec{\boldsymbol{I_C}} \cdot \vec{\boldsymbol{B}} + \vec{\boldsymbol{S}} \cdot \boldsymbol{A} \cdot \vec{\boldsymbol{I_C}},
\end{equation}
where $H_{P1}$ is shown in equation 1 of the main text and the same hyperfine tensor $\boldsymbol{A}$ from equation \ref{NV-C13} describes the interaction between the P1 electron spin and $^{13}$C spin. Approximating the spins as point dipoles, separated by a distance $r$, the hyperfine interaction tensor between the electron spin and $^{13}$C in both systems is given by:
\begin{equation}
\boldsymbol{A} = \frac{\upmu_0 \gamma_e \gamma_{\textrm{13C}} \hbar}{4 \pi r^3}(1 - 3 \hat{\boldsymbol{r}} \hat{\boldsymbol{r}}),
\end{equation}
where $\upmu_0$ is the vacuum permittivity and $\hat{\boldsymbol{r}}$ is a unit vector along the axis connecting the central electron spin to the $^{13}$C spin.

\section*{Appendix C}
\vspace{-0.3cm} \textbf{Disjoint Cluster Simulations.} We use the disjoint cluster model \cite{maze_electron_2008} to simulate the evolution of the P1 spins in the presence of a $^{13}$C spin bath. In this method,  a bath of $N$ $^{13}$Cs is randomly distributed around a central P1 spin. The ensemble of $N$ interacting spins is sorted into $n$ groups $\{G_1, G_2, ... , G_n\}$ containing up to $g$ individual spins. The spins are grouped such that there are strong interactions between the elements of each group and negligible interactions between elements in different groups. We consider a single P1 center interacting with a group of up to $g$ $^{13}$C nuclear spins. From Eq. \ref{P1-C13}, the Hamiltonian for this system is given by:
\begin{equation}
	H = H_{P1} + \gamma_{\textrm{13C}}\sum_{i = g} \vec{\boldsymbol{I}}_{C, i} \cdot \vec{\boldsymbol{B}} + \sum_{i = g} \vec{\boldsymbol{S}} \cdot \boldsymbol{A}_i \cdot \vec{\boldsymbol{I}}_{C, i}
\end{equation}

The density matrix of this system is $\rho = \rho_e \otimes \rho_N$, where $\rho_e$ is the density matrix of the P1 electron spin and $\rho_N = 1/2^{g} \mathbb{1}^{g}$ is the density matrix for the $^{13}$C spin bath, which is assumed to be thermally-populated. Following a spin-echo sequence, the density matrix of the system is:
\begin{equation}
\rho_f = U_{se, \tau} \, \rho_{i} \, U_{se, \tau}^{\dagger}, 
\end{equation}
where $U_{se, \tau} = U_{\pi/2} U(\tau) U_{\pi} U(\tau) U_{\pi/2}$ is a unitary operator denoting a spin-echo sequence applied to the P1 electron spin with a free evolution time $\tau$ and $\rho_{i}$ is the density matrix describing the initial state of the system. The operators, $U_{\pi/2}$ and $U_{\pi}$, represent $\pi/2$ and $\pi$ rf pulses. For convenience, the P1 electron spin is initialized in the $\ket{m_S = +1/2}$ state and, therefore, the initial density matrix is set as:
\begin{equation}
\rho_i = \ket{+1/2} \bra{+1/2} \otimes \rho_N. 
\end{equation}
The normalized spin-echo signal for a single group, $G$, corresponds to the population of the $\ket{m_s = +1/2}$ state, which is calculated as:
\begin{equation}
S_G  = 2\textrm{Tr} \, [\mathcal{P}_{+} \rho_{f}] - 1,
\end{equation}
where $\mathcal{P}_{+}$ is the projection operator onto the $\ket{m_s = +1/2}$ P1 electron spin state. The total spin-echo signal, $S_T$ for a single P1 center interacting with a bath of nuclear spins is then:
\begin{equation}
S_T = \prod_{G} S_G.
\end{equation}
We simulate ensemble averaging, to match the conditions in the experiment, by computing the average signal: 
\begin{equation}
\label{avesig}
S_{\textrm{ave}} = 1/N_{\textrm{B}} \sum_{N_{\textrm{B}}} S_T
\end{equation}
for $N_{\textrm{B}}$ distributions of $^{13}$C spin baths. The average spin-echo signal obtained from Eq. \ref{avesig} is calculated for a range of free evolution times, $\tau$, to simulate the time-dependent P1 spin-echo signal.

\bibliographystyle{apsrev4-2}

\begin{thebibliography}{62}%
\makeatletter
\providecommand \@ifxundefined [1]{%
 \@ifx{#1\undefined}
}%
\providecommand \@ifnum [1]{%
 \ifnum #1\expandafter \@firstoftwo
 \else \expandafter \@secondoftwo
 \fi
}%
\providecommand \@ifx [1]{%
 \ifx #1\expandafter \@firstoftwo
 \else \expandafter \@secondoftwo
 \fi
}%
\providecommand \natexlab [1]{#1}%
\providecommand \enquote  [1]{``#1''}%
\providecommand \bibnamefont  [1]{#1}%
\providecommand \bibfnamefont [1]{#1}%
\providecommand \citenamefont [1]{#1}%
\providecommand \href@noop [0]{\@secondoftwo}%
\providecommand \href [0]{\begingroup \@sanitize@url \@href}%
\providecommand \@href[1]{\@@startlink{#1}\@@href}%
\providecommand \@@href[1]{\endgroup#1\@@endlink}%
\providecommand \@sanitize@url [0]{\catcode `\\12\catcode `\$12\catcode
  `\&12\catcode `\#12\catcode `\^12\catcode `\_12\catcode `\%12\relax}%
\providecommand \@@startlink[1]{}%
\providecommand \@@endlink[0]{}%
\providecommand \url  [0]{\begingroup\@sanitize@url \@url }%
\providecommand \@url [1]{\endgroup\@href {#1}{\urlprefix }}%
\providecommand \urlprefix  [0]{URL }%
\providecommand \Eprint [0]{\href }%
\providecommand \doibase [0]{https://doi.org/}%
\providecommand \selectlanguage [0]{\@gobble}%
\providecommand \bibinfo  [0]{\@secondoftwo}%
\providecommand \bibfield  [0]{\@secondoftwo}%
\providecommand \translation [1]{[#1]}%
\providecommand \BibitemOpen [0]{}%
\providecommand \bibitemStop [0]{}%
\providecommand \bibitemNoStop [0]{.\EOS\space}%
\providecommand \EOS [0]{\spacefactor3000\relax}%
\providecommand \BibitemShut  [1]{\csname bibitem#1\endcsname}%
\let\auto@bib@innerbib\@empty
\bibitem [{\citenamefont {de~Lange}\ \emph {et~al.}(2010)\citenamefont
  {de~Lange}, \citenamefont {Wang}, \citenamefont {Ristè}, \citenamefont
  {Dobrovitski},\ and\ \citenamefont {Hanson}}]{de_lange_universal_2010}%
  \BibitemOpen
  \bibfield  {author} {\bibinfo {author} {\bibfnamefont {G.}~\bibnamefont
  {de~Lange}}, \bibinfo {author} {\bibfnamefont {Z.~H.}\ \bibnamefont {Wang}},
  \bibinfo {author} {\bibfnamefont {D.}~\bibnamefont {Ristè}}, \bibinfo
  {author} {\bibfnamefont {V.~V.}\ \bibnamefont {Dobrovitski}},\ and\ \bibinfo
  {author} {\bibfnamefont {R.}~\bibnamefont {Hanson}},\ }\href
  {https://doi.org/10.1126/science.1192739} {\bibfield  {journal} {\bibinfo
  {journal} {Science}\ }\textbf {\bibinfo {volume} {330}},\ \bibinfo {pages}
  {60} (\bibinfo {year} {2010})}\BibitemShut {NoStop}%
\bibitem [{\citenamefont {de~Lange}\ \emph {et~al.}(2012)\citenamefont
  {de~Lange}, \citenamefont {van~der Sar}, \citenamefont {Blok}, \citenamefont
  {Wang}, \citenamefont {Dobrovitski},\ and\ \citenamefont
  {Hanson}}]{de_lange_controlling_2012}%
  \BibitemOpen
  \bibfield  {author} {\bibinfo {author} {\bibfnamefont {G.}~\bibnamefont
  {de~Lange}}, \bibinfo {author} {\bibfnamefont {T.}~\bibnamefont {van~der
  Sar}}, \bibinfo {author} {\bibfnamefont {M.}~\bibnamefont {Blok}}, \bibinfo
  {author} {\bibfnamefont {Z.-H.}\ \bibnamefont {Wang}}, \bibinfo {author}
  {\bibfnamefont {V.}~\bibnamefont {Dobrovitski}},\ and\ \bibinfo {author}
  {\bibfnamefont {R.}~\bibnamefont {Hanson}},\ }\href
  {https://doi.org/10.1038/srep00382} {\bibfield  {journal} {\bibinfo
  {journal} {Sci. Rep.}\ }\textbf {\bibinfo {volume} {2}},\ \bibinfo {pages}
  {382} (\bibinfo {year} {2012})}\BibitemShut {NoStop}%
\bibitem [{\citenamefont {Bar-Gill}\ \emph {et~al.}(2012)\citenamefont
  {Bar-Gill}, \citenamefont {Pham}, \citenamefont {Belthangady}, \citenamefont
  {Le~Sage}, \citenamefont {Cappellaro}, \citenamefont {Maze}, \citenamefont
  {Lukin}, \citenamefont {Yacoby},\ and\ \citenamefont
  {Walsworth}}]{bar-gill_suppression_2012}%
  \BibitemOpen
  \bibfield  {author} {\bibinfo {author} {\bibfnamefont {N.}~\bibnamefont
  {Bar-Gill}}, \bibinfo {author} {\bibfnamefont {L.~M.}\ \bibnamefont {Pham}},
  \bibinfo {author} {\bibfnamefont {C.}~\bibnamefont {Belthangady}}, \bibinfo
  {author} {\bibfnamefont {D.}~\bibnamefont {Le~Sage}}, \bibinfo {author}
  {\bibfnamefont {P.}~\bibnamefont {Cappellaro}}, \bibinfo {author}
  {\bibfnamefont {J.~R.}\ \bibnamefont {Maze}}, \bibinfo {author}
  {\bibfnamefont {M.~D.}\ \bibnamefont {Lukin}}, \bibinfo {author}
  {\bibfnamefont {A.}~\bibnamefont {Yacoby}},\ and\ \bibinfo {author}
  {\bibfnamefont {R.}~\bibnamefont {Walsworth}},\ }\href
  {https://doi.org/10.1038/ncomms1856} {\bibfield  {journal} {\bibinfo
  {journal} {Nat. Commun.}\ }\textbf {\bibinfo {volume} {3}},\ \bibinfo {pages}
  {858} (\bibinfo {year} {2012})}\BibitemShut {NoStop}%
\bibitem [{\citenamefont {Tyryshkin}\ \emph {et~al.}(2012)\citenamefont
  {Tyryshkin}, \citenamefont {Tojo}, \citenamefont {Morton}, \citenamefont
  {Riemann}, \citenamefont {Abrosimov}, \citenamefont {Becker}, \citenamefont
  {Pohl}, \citenamefont {Schenkel}, \citenamefont {Thewalt}, \citenamefont
  {Itoh},\ and\ \citenamefont {Lyon}}]{tyryshkin_electron_2012}%
  \BibitemOpen
  \bibfield  {author} {\bibinfo {author} {\bibfnamefont {A.~M.}\ \bibnamefont
  {Tyryshkin}}, \bibinfo {author} {\bibfnamefont {S.}~\bibnamefont {Tojo}},
  \bibinfo {author} {\bibfnamefont {J.~J.~L.}\ \bibnamefont {Morton}}, \bibinfo
  {author} {\bibfnamefont {H.}~\bibnamefont {Riemann}}, \bibinfo {author}
  {\bibfnamefont {N.~V.}\ \bibnamefont {Abrosimov}}, \bibinfo {author}
  {\bibfnamefont {P.}~\bibnamefont {Becker}}, \bibinfo {author} {\bibfnamefont
  {H.-J.}\ \bibnamefont {Pohl}}, \bibinfo {author} {\bibfnamefont
  {T.}~\bibnamefont {Schenkel}}, \bibinfo {author} {\bibfnamefont {M.~L.~W.}\
  \bibnamefont {Thewalt}}, \bibinfo {author} {\bibfnamefont {K.~M.}\
  \bibnamefont {Itoh}},\ and\ \bibinfo {author} {\bibfnamefont {S.~A.}\
  \bibnamefont {Lyon}},\ }\href {https://doi.org/10.1038/nmat3182} {\bibfield
  {journal} {\bibinfo  {journal} {Nat. Mater.}\ }\textbf {\bibinfo {volume}
  {11}},\ \bibinfo {pages} {143} (\bibinfo {year} {2012})}\BibitemShut
  {NoStop}%
\bibitem [{\citenamefont {Bauch}\ \emph {et~al.}(2018)\citenamefont {Bauch},
  \citenamefont {Hart}, \citenamefont {Schloss}, \citenamefont {Turner},
  \citenamefont {Barry}, \citenamefont {Kehayias}, \citenamefont {Singh},\ and\
  \citenamefont {Walsworth}}]{bauch_ultralong_2018}%
  \BibitemOpen
  \bibfield  {author} {\bibinfo {author} {\bibfnamefont {E.}~\bibnamefont
  {Bauch}}, \bibinfo {author} {\bibfnamefont {C.~A.}\ \bibnamefont {Hart}},
  \bibinfo {author} {\bibfnamefont {J.~M.}\ \bibnamefont {Schloss}}, \bibinfo
  {author} {\bibfnamefont {M.~J.}\ \bibnamefont {Turner}}, \bibinfo {author}
  {\bibfnamefont {J.~F.}\ \bibnamefont {Barry}}, \bibinfo {author}
  {\bibfnamefont {P.}~\bibnamefont {Kehayias}}, \bibinfo {author}
  {\bibfnamefont {S.}~\bibnamefont {Singh}},\ and\ \bibinfo {author}
  {\bibfnamefont {R.~L.}\ \bibnamefont {Walsworth}},\ }\href
  {https://doi.org/10.1103/PhysRevX.8.031025} {\bibfield  {journal} {\bibinfo
  {journal} {Phys. Rev. X.}\ }\textbf {\bibinfo {volume} {8}},\ \bibinfo
  {pages} {031025} (\bibinfo {year} {2018})}\BibitemShut {NoStop}%
\bibitem [{\citenamefont {Cai}\ \emph {et~al.}(2013)\citenamefont {Cai},
  \citenamefont {Retzker}, \citenamefont {Jelezko},\ and\ \citenamefont
  {Plenio}}]{cai_large-scale_2013}%
  \BibitemOpen
  \bibfield  {author} {\bibinfo {author} {\bibfnamefont {J.}~\bibnamefont
  {Cai}}, \bibinfo {author} {\bibfnamefont {A.}~\bibnamefont {Retzker}},
  \bibinfo {author} {\bibfnamefont {F.}~\bibnamefont {Jelezko}},\ and\ \bibinfo
  {author} {\bibfnamefont {M.~B.}\ \bibnamefont {Plenio}},\ }\href
  {https://doi.org/10.1038/nphys2519} {\bibfield  {journal} {\bibinfo
  {journal} {Nat. Phys.}\ }\textbf {\bibinfo {volume} {9}},\ \bibinfo {pages}
  {168} (\bibinfo {year} {2013})}\BibitemShut {NoStop}%
\bibitem [{\citenamefont {Unden}\ \emph {et~al.}(2018)\citenamefont {Unden},
  \citenamefont {Tomek}, \citenamefont {Weggler}, \citenamefont {Frank},
  \citenamefont {London}, \citenamefont {Zopes}, \citenamefont {Degen},
  \citenamefont {Raatz}, \citenamefont {Meijer}, \citenamefont {Watanabe},
  \citenamefont {Itoh}, \citenamefont {Plenio}, \citenamefont {Naydenov},\ and\
  \citenamefont {Jelezko}}]{unden_coherent_2018}%
  \BibitemOpen
  \bibfield  {author} {\bibinfo {author} {\bibfnamefont {T.}~\bibnamefont
  {Unden}}, \bibinfo {author} {\bibfnamefont {N.}~\bibnamefont {Tomek}},
  \bibinfo {author} {\bibfnamefont {T.}~\bibnamefont {Weggler}}, \bibinfo
  {author} {\bibfnamefont {F.}~\bibnamefont {Frank}}, \bibinfo {author}
  {\bibfnamefont {P.}~\bibnamefont {London}}, \bibinfo {author} {\bibfnamefont
  {J.}~\bibnamefont {Zopes}}, \bibinfo {author} {\bibfnamefont
  {C.}~\bibnamefont {Degen}}, \bibinfo {author} {\bibfnamefont
  {N.}~\bibnamefont {Raatz}}, \bibinfo {author} {\bibfnamefont
  {J.}~\bibnamefont {Meijer}}, \bibinfo {author} {\bibfnamefont
  {H.}~\bibnamefont {Watanabe}}, \bibinfo {author} {\bibfnamefont {K.~M.}\
  \bibnamefont {Itoh}}, \bibinfo {author} {\bibfnamefont {M.~B.}\ \bibnamefont
  {Plenio}}, \bibinfo {author} {\bibfnamefont {B.}~\bibnamefont {Naydenov}},\
  and\ \bibinfo {author} {\bibfnamefont {F.}~\bibnamefont {Jelezko}},\ }\href
  {https://doi.org/10.1038/s41534-018-0089-8} {\bibfield  {journal} {\bibinfo
  {journal} {npj Quantum Inf.}\ }\textbf {\bibinfo {volume} {4}},\ \bibinfo
  {pages} {1} (\bibinfo {year} {2018})}\BibitemShut {NoStop}%
\bibitem [{\citenamefont {Hermans}\ \emph {et~al.}(2022)\citenamefont
  {Hermans}, \citenamefont {Pompili}, \citenamefont {Beukers}, \citenamefont
  {Baier}, \citenamefont {Borregaard},\ and\ \citenamefont
  {Hanson}}]{hermans_qubit_2022}%
  \BibitemOpen
  \bibfield  {author} {\bibinfo {author} {\bibfnamefont {S.~L.~N.}\
  \bibnamefont {Hermans}}, \bibinfo {author} {\bibfnamefont {M.}~\bibnamefont
  {Pompili}}, \bibinfo {author} {\bibfnamefont {H.~K.~C.}\ \bibnamefont
  {Beukers}}, \bibinfo {author} {\bibfnamefont {S.}~\bibnamefont {Baier}},
  \bibinfo {author} {\bibfnamefont {J.}~\bibnamefont {Borregaard}},\ and\
  \bibinfo {author} {\bibfnamefont {R.}~\bibnamefont {Hanson}},\ }\href
  {https://doi.org/10.1038/s41586-022-04697-y} {\bibfield  {journal} {\bibinfo
  {journal} {Nature}\ }\textbf {\bibinfo {volume} {605}},\ \bibinfo {pages}
  {663} (\bibinfo {year} {2022})}\BibitemShut {NoStop}%
\bibitem [{\citenamefont {Cujia}\ \emph {et~al.}(2022)\citenamefont {Cujia},
  \citenamefont {Herb}, \citenamefont {Zopes}, \citenamefont {Abendroth},\ and\
  \citenamefont {Degen}}]{cujia_parallel_2022}%
  \BibitemOpen
  \bibfield  {author} {\bibinfo {author} {\bibfnamefont {K.~S.}\ \bibnamefont
  {Cujia}}, \bibinfo {author} {\bibfnamefont {K.}~\bibnamefont {Herb}},
  \bibinfo {author} {\bibfnamefont {J.}~\bibnamefont {Zopes}}, \bibinfo
  {author} {\bibfnamefont {J.~M.}\ \bibnamefont {Abendroth}},\ and\ \bibinfo
  {author} {\bibfnamefont {C.~L.}\ \bibnamefont {Degen}},\ }\href
  {https://doi.org/10.1038/s41467-022-28935-z} {\bibfield  {journal} {\bibinfo
  {journal} {Nat. Commun.}\ }\textbf {\bibinfo {volume} {13}},\ \bibinfo
  {pages} {1260} (\bibinfo {year} {2022})}\BibitemShut {NoStop}%
\bibitem [{\citenamefont {Doherty}\ \emph {et~al.}(2013)\citenamefont
  {Doherty}, \citenamefont {Manson}, \citenamefont {Delaney}, \citenamefont
  {Jelezko}, \citenamefont {Wrachtrup},\ and\ \citenamefont
  {Hollenberg}}]{doherty_nitrogen-vacancy_2013}%
  \BibitemOpen
  \bibfield  {author} {\bibinfo {author} {\bibfnamefont {M.~W.}\ \bibnamefont
  {Doherty}}, \bibinfo {author} {\bibfnamefont {N.~B.}\ \bibnamefont {Manson}},
  \bibinfo {author} {\bibfnamefont {P.}~\bibnamefont {Delaney}}, \bibinfo
  {author} {\bibfnamefont {F.}~\bibnamefont {Jelezko}}, \bibinfo {author}
  {\bibfnamefont {J.}~\bibnamefont {Wrachtrup}},\ and\ \bibinfo {author}
  {\bibfnamefont {L.~C.~L.}\ \bibnamefont {Hollenberg}},\ }\href
  {https://doi.org/10.1016/j.physrep.2013.02.001} {\bibfield  {journal}
  {\bibinfo  {journal} {Phys. Rep.}\ }\textbf {\bibinfo {volume} {528}},\
  \bibinfo {pages} {1} (\bibinfo {year} {2013})}\BibitemShut {NoStop}%
\bibitem [{\citenamefont {Barry}\ \emph {et~al.}(2020)\citenamefont {Barry},
  \citenamefont {Schloss}, \citenamefont {Bauch}, \citenamefont {Turner},
  \citenamefont {Hart}, \citenamefont {Pham},\ and\ \citenamefont
  {Walsworth}}]{barry_sensitivity_2020}%
  \BibitemOpen
  \bibfield  {author} {\bibinfo {author} {\bibfnamefont {J.~F.}\ \bibnamefont
  {Barry}}, \bibinfo {author} {\bibfnamefont {J.~M.}\ \bibnamefont {Schloss}},
  \bibinfo {author} {\bibfnamefont {E.}~\bibnamefont {Bauch}}, \bibinfo
  {author} {\bibfnamefont {M.~J.}\ \bibnamefont {Turner}}, \bibinfo {author}
  {\bibfnamefont {C.~A.}\ \bibnamefont {Hart}}, \bibinfo {author}
  {\bibfnamefont {L.~M.}\ \bibnamefont {Pham}},\ and\ \bibinfo {author}
  {\bibfnamefont {R.~L.}\ \bibnamefont {Walsworth}},\ }\href
  {https://doi.org/10.1103/RevModPhys.92.015004} {\bibfield  {journal}
  {\bibinfo  {journal} {Rev. Mod. Phys.}\ }\textbf {\bibinfo {volume} {92}},\
  \bibinfo {pages} {015004} (\bibinfo {year} {2020})}\BibitemShut {NoStop}%
\bibitem [{\citenamefont {Dutt}\ \emph {et~al.}(2007)\citenamefont {Dutt},
  \citenamefont {Childress}, \citenamefont {Jiang}, \citenamefont {Togan},
  \citenamefont {Maze}, \citenamefont {Jelezko}, \citenamefont {Zibrov},
  \citenamefont {Hemmer},\ and\ \citenamefont {Lukin}}]{dutt_quantum_2007}%
  \BibitemOpen
  \bibfield  {author} {\bibinfo {author} {\bibfnamefont {M.~V.~G.}\
  \bibnamefont {Dutt}}, \bibinfo {author} {\bibfnamefont {L.}~\bibnamefont
  {Childress}}, \bibinfo {author} {\bibfnamefont {L.}~\bibnamefont {Jiang}},
  \bibinfo {author} {\bibfnamefont {E.}~\bibnamefont {Togan}}, \bibinfo
  {author} {\bibfnamefont {J.}~\bibnamefont {Maze}}, \bibinfo {author}
  {\bibfnamefont {F.}~\bibnamefont {Jelezko}}, \bibinfo {author} {\bibfnamefont
  {A.~S.}\ \bibnamefont {Zibrov}}, \bibinfo {author} {\bibfnamefont {P.~R.}\
  \bibnamefont {Hemmer}},\ and\ \bibinfo {author} {\bibfnamefont {M.~D.}\
  \bibnamefont {Lukin}},\ }\href {https://doi.org/10.1126/science.1139831}
  {\bibfield  {journal} {\bibinfo  {journal} {Science}\ }\textbf {\bibinfo
  {volume} {316}},\ \bibinfo {pages} {1312} (\bibinfo {year}
  {2007})}\BibitemShut {NoStop}%
\bibitem [{\citenamefont {Neumann}\ \emph {et~al.}(2008)\citenamefont
  {Neumann}, \citenamefont {Mizuochi}, \citenamefont {Rempp}, \citenamefont
  {Hemmer}, \citenamefont {Watanabe}, \citenamefont {Yamasaki}, \citenamefont
  {Jacques}, \citenamefont {Gaebel}, \citenamefont {Jelezko},\ and\
  \citenamefont {Wrachtrup}}]{neumann_multipartite_2008}%
  \BibitemOpen
  \bibfield  {author} {\bibinfo {author} {\bibfnamefont {P.}~\bibnamefont
  {Neumann}}, \bibinfo {author} {\bibfnamefont {N.}~\bibnamefont {Mizuochi}},
  \bibinfo {author} {\bibfnamefont {F.}~\bibnamefont {Rempp}}, \bibinfo
  {author} {\bibfnamefont {P.}~\bibnamefont {Hemmer}}, \bibinfo {author}
  {\bibfnamefont {H.}~\bibnamefont {Watanabe}}, \bibinfo {author}
  {\bibfnamefont {S.}~\bibnamefont {Yamasaki}}, \bibinfo {author}
  {\bibfnamefont {V.}~\bibnamefont {Jacques}}, \bibinfo {author} {\bibfnamefont
  {T.}~\bibnamefont {Gaebel}}, \bibinfo {author} {\bibfnamefont
  {F.}~\bibnamefont {Jelezko}},\ and\ \bibinfo {author} {\bibfnamefont
  {J.}~\bibnamefont {Wrachtrup}},\ }\href
  {https://doi.org/10.1126/science.1157233} {\bibfield  {journal} {\bibinfo
  {journal} {Science}\ }\textbf {\bibinfo {volume} {320}},\ \bibinfo {pages}
  {1326} (\bibinfo {year} {2008})}\BibitemShut {NoStop}%
\bibitem [{\citenamefont {Maurer}\ \emph {et~al.}(2012)\citenamefont {Maurer},
  \citenamefont {Kucsko}, \citenamefont {Latta}, \citenamefont {Jiang},
  \citenamefont {Yao}, \citenamefont {Bennett}, \citenamefont {Pastawski},
  \citenamefont {Hunger}, \citenamefont {Chisholm}, \citenamefont {Markham},
  \citenamefont {Twitchen}, \citenamefont {Cirac},\ and\ \citenamefont
  {Lukin}}]{maurer_room-temperature_2012}%
  \BibitemOpen
  \bibfield  {author} {\bibinfo {author} {\bibfnamefont {P.~C.}\ \bibnamefont
  {Maurer}}, \bibinfo {author} {\bibfnamefont {G.}~\bibnamefont {Kucsko}},
  \bibinfo {author} {\bibfnamefont {C.}~\bibnamefont {Latta}}, \bibinfo
  {author} {\bibfnamefont {L.}~\bibnamefont {Jiang}}, \bibinfo {author}
  {\bibfnamefont {N.~Y.}\ \bibnamefont {Yao}}, \bibinfo {author} {\bibfnamefont
  {S.~D.}\ \bibnamefont {Bennett}}, \bibinfo {author} {\bibfnamefont
  {F.}~\bibnamefont {Pastawski}}, \bibinfo {author} {\bibfnamefont
  {D.}~\bibnamefont {Hunger}}, \bibinfo {author} {\bibfnamefont
  {N.}~\bibnamefont {Chisholm}}, \bibinfo {author} {\bibfnamefont
  {M.}~\bibnamefont {Markham}}, \bibinfo {author} {\bibfnamefont {D.~J.}\
  \bibnamefont {Twitchen}}, \bibinfo {author} {\bibfnamefont {J.~I.}\
  \bibnamefont {Cirac}},\ and\ \bibinfo {author} {\bibfnamefont {M.~D.}\
  \bibnamefont {Lukin}},\ }\href {https://doi.org/10.1126/science.1220513}
  {\bibfield  {journal} {\bibinfo  {journal} {Science}\ }\textbf {\bibinfo
  {volume} {336}},\ \bibinfo {pages} {1283} (\bibinfo {year}
  {2012})}\BibitemShut {NoStop}%
\bibitem [{\citenamefont {Taminiau}\ \emph {et~al.}(2014)\citenamefont
  {Taminiau}, \citenamefont {Cramer}, \citenamefont {van~der Sar},
  \citenamefont {Dobrovitski},\ and\ \citenamefont
  {Hanson}}]{taminiau_universal_2014}%
  \BibitemOpen
  \bibfield  {author} {\bibinfo {author} {\bibfnamefont {T.~H.}\ \bibnamefont
  {Taminiau}}, \bibinfo {author} {\bibfnamefont {J.}~\bibnamefont {Cramer}},
  \bibinfo {author} {\bibfnamefont {T.}~\bibnamefont {van~der Sar}}, \bibinfo
  {author} {\bibfnamefont {V.~V.}\ \bibnamefont {Dobrovitski}},\ and\ \bibinfo
  {author} {\bibfnamefont {R.}~\bibnamefont {Hanson}},\ }\href
  {https://doi.org/10.1038/nnano.2014.2} {\bibfield  {journal} {\bibinfo
  {journal} {Nat. Nanotechnol.}\ }\textbf {\bibinfo {volume} {9}},\ \bibinfo
  {pages} {171} (\bibinfo {year} {2014})}\BibitemShut {NoStop}%
\bibitem [{\citenamefont {Abobeih}\ \emph {et~al.}(2019)\citenamefont
  {Abobeih}, \citenamefont {Randall}, \citenamefont {Bradley}, \citenamefont
  {Bartling}, \citenamefont {Bakker}, \citenamefont {Degen}, \citenamefont
  {Markham}, \citenamefont {Twitchen},\ and\ \citenamefont
  {Taminiau}}]{abobeih_atomic-scale_2019}%
  \BibitemOpen
  \bibfield  {author} {\bibinfo {author} {\bibfnamefont {M.~H.}\ \bibnamefont
  {Abobeih}}, \bibinfo {author} {\bibfnamefont {J.}~\bibnamefont {Randall}},
  \bibinfo {author} {\bibfnamefont {C.~E.}\ \bibnamefont {Bradley}}, \bibinfo
  {author} {\bibfnamefont {H.~P.}\ \bibnamefont {Bartling}}, \bibinfo {author}
  {\bibfnamefont {M.~A.}\ \bibnamefont {Bakker}}, \bibinfo {author}
  {\bibfnamefont {M.~J.}\ \bibnamefont {Degen}}, \bibinfo {author}
  {\bibfnamefont {M.}~\bibnamefont {Markham}}, \bibinfo {author} {\bibfnamefont
  {D.~J.}\ \bibnamefont {Twitchen}},\ and\ \bibinfo {author} {\bibfnamefont
  {T.~H.}\ \bibnamefont {Taminiau}},\ }\href
  {https://doi.org/10.1038/s41586-019-1834-7} {\bibfield  {journal} {\bibinfo
  {journal} {Nature}\ }\textbf {\bibinfo {volume} {576}},\ \bibinfo {pages}
  {411} (\bibinfo {year} {2019})}\BibitemShut {NoStop}%
\bibitem [{\citenamefont {Abobeih}\ \emph {et~al.}(2022)\citenamefont
  {Abobeih}, \citenamefont {Wang}, \citenamefont {Randall}, \citenamefont
  {Loenen}, \citenamefont {Bradley}, \citenamefont {Markham}, \citenamefont
  {Twitchen}, \citenamefont {Terhal},\ and\ \citenamefont
  {Taminiau}}]{abobeih_fault-tolerant_2022}%
  \BibitemOpen
  \bibfield  {author} {\bibinfo {author} {\bibfnamefont {M.~H.}\ \bibnamefont
  {Abobeih}}, \bibinfo {author} {\bibfnamefont {Y.}~\bibnamefont {Wang}},
  \bibinfo {author} {\bibfnamefont {J.}~\bibnamefont {Randall}}, \bibinfo
  {author} {\bibfnamefont {S.~J.~H.}\ \bibnamefont {Loenen}}, \bibinfo {author}
  {\bibfnamefont {C.~E.}\ \bibnamefont {Bradley}}, \bibinfo {author}
  {\bibfnamefont {M.}~\bibnamefont {Markham}}, \bibinfo {author} {\bibfnamefont
  {D.~J.}\ \bibnamefont {Twitchen}}, \bibinfo {author} {\bibfnamefont {B.~M.}\
  \bibnamefont {Terhal}},\ and\ \bibinfo {author} {\bibfnamefont {T.~H.}\
  \bibnamefont {Taminiau}},\ }\href
  {https://doi.org/10.1038/s41586-022-04819-6} {\bibfield  {journal} {\bibinfo
  {journal} {Nature}\ }\textbf {\bibinfo {volume} {606}},\ \bibinfo {pages}
  {884} (\bibinfo {year} {2022})}\BibitemShut {NoStop}%
\bibitem [{\citenamefont {van~de Stolpe}\ \emph {et~al.}(2023)\citenamefont
  {van~de Stolpe}, \citenamefont {Kwiatkowski}, \citenamefont {Bradley},
  \citenamefont {Randall}, \citenamefont {Breitweiser}, \citenamefont
  {Bassett}, \citenamefont {Markham}, \citenamefont {Twitchen},\ and\
  \citenamefont {Taminiau}}]{van_de_stolpe_mapping_2023}%
  \BibitemOpen
  \bibfield  {author} {\bibinfo {author} {\bibfnamefont {G.~L.}\ \bibnamefont
  {van~de Stolpe}}, \bibinfo {author} {\bibfnamefont {D.~P.}\ \bibnamefont
  {Kwiatkowski}}, \bibinfo {author} {\bibfnamefont {C.~E.}\ \bibnamefont
  {Bradley}}, \bibinfo {author} {\bibfnamefont {J.}~\bibnamefont {Randall}},
  \bibinfo {author} {\bibfnamefont {S.~A.}\ \bibnamefont {Breitweiser}},
  \bibinfo {author} {\bibfnamefont {L.~C.}\ \bibnamefont {Bassett}}, \bibinfo
  {author} {\bibfnamefont {M.}~\bibnamefont {Markham}}, \bibinfo {author}
  {\bibfnamefont {D.~J.}\ \bibnamefont {Twitchen}},\ and\ \bibinfo {author}
  {\bibfnamefont {T.~H.}\ \bibnamefont {Taminiau}},\ }\Eprint
  {https://arxiv.org/abs/2307.06939} {arXiv:2307.06939 [quant-ph]}  (\bibinfo
  {year} {2023})\BibitemShut {NoStop}%
\bibitem [{\citenamefont {Laraoui}\ and\ \citenamefont
  {Meriles}(2013)}]{laraoui_approach_2013}%
  \BibitemOpen
  \bibfield  {author} {\bibinfo {author} {\bibfnamefont {A.}~\bibnamefont
  {Laraoui}}\ and\ \bibinfo {author} {\bibfnamefont {C.~A.}\ \bibnamefont
  {Meriles}},\ }\href {https://doi.org/10.1021/nn400239n} {\bibfield  {journal}
  {\bibinfo  {journal} {ACS Nano}\ }\textbf {\bibinfo {volume} {7}},\ \bibinfo
  {pages} {3403} (\bibinfo {year} {2013})}\BibitemShut {NoStop}%
\bibitem [{\citenamefont {Belthangady}\ \emph {et~al.}(2013)\citenamefont
  {Belthangady}, \citenamefont {Bar-Gill}, \citenamefont {Pham}, \citenamefont
  {Arai}, \citenamefont {Le~Sage}, \citenamefont {Cappellaro},\ and\
  \citenamefont {Walsworth}}]{belthangady_dressed-state_2013}%
  \BibitemOpen
  \bibfield  {author} {\bibinfo {author} {\bibfnamefont {C.}~\bibnamefont
  {Belthangady}}, \bibinfo {author} {\bibfnamefont {N.}~\bibnamefont
  {Bar-Gill}}, \bibinfo {author} {\bibfnamefont {L.~M.}\ \bibnamefont {Pham}},
  \bibinfo {author} {\bibfnamefont {K.}~\bibnamefont {Arai}}, \bibinfo {author}
  {\bibfnamefont {D.}~\bibnamefont {Le~Sage}}, \bibinfo {author} {\bibfnamefont
  {P.}~\bibnamefont {Cappellaro}},\ and\ \bibinfo {author} {\bibfnamefont
  {R.~L.}\ \bibnamefont {Walsworth}},\ }\href
  {https://doi.org/10.1103/PhysRevLett.110.157601} {\bibfield  {journal}
  {\bibinfo  {journal} {Phys. Rev. Lett.}\ }\textbf {\bibinfo {volume} {110}},\
  \bibinfo {pages} {157601} (\bibinfo {year} {2013})}\BibitemShut {NoStop}%
\bibitem [{\citenamefont {Degen}\ \emph {et~al.}(2021)\citenamefont {Degen},
  \citenamefont {Loenen}, \citenamefont {Bartling}, \citenamefont {Bradley},
  \citenamefont {Meinsma}, \citenamefont {Markham}, \citenamefont {Twitchen},\
  and\ \citenamefont {Taminiau}}]{degen_entanglement_2021}%
  \BibitemOpen
  \bibfield  {author} {\bibinfo {author} {\bibfnamefont {M.~J.}\ \bibnamefont
  {Degen}}, \bibinfo {author} {\bibfnamefont {S.~J.~H.}\ \bibnamefont
  {Loenen}}, \bibinfo {author} {\bibfnamefont {H.~P.}\ \bibnamefont
  {Bartling}}, \bibinfo {author} {\bibfnamefont {C.~E.}\ \bibnamefont
  {Bradley}}, \bibinfo {author} {\bibfnamefont {A.~L.}\ \bibnamefont
  {Meinsma}}, \bibinfo {author} {\bibfnamefont {M.}~\bibnamefont {Markham}},
  \bibinfo {author} {\bibfnamefont {D.~J.}\ \bibnamefont {Twitchen}},\ and\
  \bibinfo {author} {\bibfnamefont {T.~H.}\ \bibnamefont {Taminiau}},\ }\href
  {https://doi.org/10.1038/s41467-021-23454-9} {\bibfield  {journal} {\bibinfo
  {journal} {Nat. Commun.}\ }\textbf {\bibinfo {volume} {12}},\ \bibinfo
  {pages} {3470} (\bibinfo {year} {2021})}\BibitemShut {NoStop}%
\bibitem [{\citenamefont {Goldblatt}\ \emph {et~al.}(2022)\citenamefont
  {Goldblatt}, \citenamefont {Martin},\ and\ \citenamefont
  {Wood}}]{goldblatt_tunable_2022}%
  \BibitemOpen
  \bibfield  {author} {\bibinfo {author} {\bibfnamefont {R.~M.}\ \bibnamefont
  {Goldblatt}}, \bibinfo {author} {\bibfnamefont {A.~M.}\ \bibnamefont
  {Martin}},\ and\ \bibinfo {author} {\bibfnamefont {A.~A.}\ \bibnamefont
  {Wood}},\ }\href {https://doi.org/10.1103/PhysRevB.105.L020405} {\bibfield
  {journal} {\bibinfo  {journal} {Phys. Rev. B.}\ }\textbf {\bibinfo {volume}
  {105}},\ \bibinfo {pages} {L020405} (\bibinfo {year} {2022})}\BibitemShut
  {NoStop}%
\bibitem [{\citenamefont {Goldstein}\ \emph {et~al.}(2011)\citenamefont
  {Goldstein}, \citenamefont {Cappellaro}, \citenamefont {Maze}, \citenamefont
  {Hodges}, \citenamefont {Jiang}, \citenamefont {Sørensen},\ and\
  \citenamefont {Lukin}}]{goldstein_environment-assisted_2011}%
  \BibitemOpen
  \bibfield  {author} {\bibinfo {author} {\bibfnamefont {G.}~\bibnamefont
  {Goldstein}}, \bibinfo {author} {\bibfnamefont {P.}~\bibnamefont
  {Cappellaro}}, \bibinfo {author} {\bibfnamefont {J.~R.}\ \bibnamefont
  {Maze}}, \bibinfo {author} {\bibfnamefont {J.~S.}\ \bibnamefont {Hodges}},
  \bibinfo {author} {\bibfnamefont {L.}~\bibnamefont {Jiang}}, \bibinfo
  {author} {\bibfnamefont {A.~S.}\ \bibnamefont {Sørensen}},\ and\ \bibinfo
  {author} {\bibfnamefont {M.~D.}\ \bibnamefont {Lukin}},\ }\href
  {https://doi.org/10.1103/PhysRevLett.106.140502} {\bibfield  {journal}
  {\bibinfo  {journal} {Phys. Rev. Lett.}\ }\textbf {\bibinfo {volume} {106}},\
  \bibinfo {pages} {140502} (\bibinfo {year} {2011})}\BibitemShut {NoStop}%
\bibitem [{\citenamefont {Knowles}\ \emph {et~al.}(2016)\citenamefont
  {Knowles}, \citenamefont {Kara},\ and\ \citenamefont
  {Atatüre}}]{knowles_demonstration_2016}%
  \BibitemOpen
  \bibfield  {author} {\bibinfo {author} {\bibfnamefont {H.~S.}\ \bibnamefont
  {Knowles}}, \bibinfo {author} {\bibfnamefont {D.~M.}\ \bibnamefont {Kara}},\
  and\ \bibinfo {author} {\bibfnamefont {M.}~\bibnamefont {Atatüre}},\ }\href
  {https://doi.org/10.1103/PhysRevLett.117.100802} {\bibfield  {journal}
  {\bibinfo  {journal} {Phys. Rev. Lett.}\ }\textbf {\bibinfo {volume} {117}},\
  \bibinfo {pages} {100802} (\bibinfo {year} {2016})}\BibitemShut {NoStop}%
\bibitem [{\citenamefont {Cooper}\ \emph {et~al.}(2019)\citenamefont {Cooper},
  \citenamefont {Sun}, \citenamefont {Jaskula},\ and\ \citenamefont
  {Cappellaro}}]{cooper_environment-assisted_2019}%
  \BibitemOpen
  \bibfield  {author} {\bibinfo {author} {\bibfnamefont {A.}~\bibnamefont
  {Cooper}}, \bibinfo {author} {\bibfnamefont {W.~K.~C.}\ \bibnamefont {Sun}},
  \bibinfo {author} {\bibfnamefont {J.-C.}\ \bibnamefont {Jaskula}},\ and\
  \bibinfo {author} {\bibfnamefont {P.}~\bibnamefont {Cappellaro}},\ }\href
  {https://doi.org/10.1103/PhysRevApplied.12.044047} {\bibfield  {journal}
  {\bibinfo  {journal} {Phys. Rev. Appl.}\ }\textbf {\bibinfo {volume} {12}},\
  \bibinfo {pages} {044047} (\bibinfo {year} {2019})}\BibitemShut {NoStop}%
\bibitem [{\citenamefont {Wunderlich}\ \emph {et~al.}(2017)\citenamefont
  {Wunderlich}, \citenamefont {Kohlrautz}, \citenamefont {Abel}, \citenamefont
  {Haase},\ and\ \citenamefont {Meijer}}]{wunderlich_optically_2017}%
  \BibitemOpen
  \bibfield  {author} {\bibinfo {author} {\bibfnamefont {R.}~\bibnamefont
  {Wunderlich}}, \bibinfo {author} {\bibfnamefont {J.}~\bibnamefont
  {Kohlrautz}}, \bibinfo {author} {\bibfnamefont {B.}~\bibnamefont {Abel}},
  \bibinfo {author} {\bibfnamefont {J.}~\bibnamefont {Haase}},\ and\ \bibinfo
  {author} {\bibfnamefont {J.}~\bibnamefont {Meijer}},\ }\href
  {https://doi.org/10.1103/PhysRevB.96.220407} {\bibfield  {journal} {\bibinfo
  {journal} {Phys. Rev. B}\ }\textbf {\bibinfo {volume} {96}},\ \bibinfo
  {pages} {220407} (\bibinfo {year} {2017})}\BibitemShut {NoStop}%
\bibitem [{\citenamefont {Pagliero}\ \emph {et~al.}(2018)\citenamefont
  {Pagliero}, \citenamefont {Rao}, \citenamefont {Zangara}, \citenamefont
  {Dhomkar}, \citenamefont {Wong}, \citenamefont {Abril}, \citenamefont
  {Aslam}, \citenamefont {Parker}, \citenamefont {King}, \citenamefont
  {Avalos}, \citenamefont {Ajoy}, \citenamefont {Wrachtrup}, \citenamefont
  {Pines},\ and\ \citenamefont {Meriles}}]{pagliero_multispin-assisted_2018}%
  \BibitemOpen
  \bibfield  {author} {\bibinfo {author} {\bibfnamefont {D.}~\bibnamefont
  {Pagliero}}, \bibinfo {author} {\bibfnamefont {K.~R.~K.}\ \bibnamefont
  {Rao}}, \bibinfo {author} {\bibfnamefont {P.~R.}\ \bibnamefont {Zangara}},
  \bibinfo {author} {\bibfnamefont {S.}~\bibnamefont {Dhomkar}}, \bibinfo
  {author} {\bibfnamefont {H.~H.}\ \bibnamefont {Wong}}, \bibinfo {author}
  {\bibfnamefont {A.}~\bibnamefont {Abril}}, \bibinfo {author} {\bibfnamefont
  {N.}~\bibnamefont {Aslam}}, \bibinfo {author} {\bibfnamefont
  {A.}~\bibnamefont {Parker}}, \bibinfo {author} {\bibfnamefont
  {J.}~\bibnamefont {King}}, \bibinfo {author} {\bibfnamefont {C.~E.}\
  \bibnamefont {Avalos}}, \bibinfo {author} {\bibfnamefont {A.}~\bibnamefont
  {Ajoy}}, \bibinfo {author} {\bibfnamefont {J.}~\bibnamefont {Wrachtrup}},
  \bibinfo {author} {\bibfnamefont {A.}~\bibnamefont {Pines}},\ and\ \bibinfo
  {author} {\bibfnamefont {C.~A.}\ \bibnamefont {Meriles}},\ }\href
  {https://doi.org/10.1103/PhysRevB.97.024422} {\bibfield  {journal} {\bibinfo
  {journal} {Phys. Rev. B.}\ }\textbf {\bibinfo {volume} {97}},\ \bibinfo
  {pages} {024422} (\bibinfo {year} {2018})}\BibitemShut {NoStop}%
\bibitem [{\citenamefont {Zangara}\ \emph {et~al.}(2019)\citenamefont
  {Zangara}, \citenamefont {Wood}, \citenamefont {Doherty},\ and\ \citenamefont
  {Meriles}}]{zangara_mechanical_2019}%
  \BibitemOpen
  \bibfield  {author} {\bibinfo {author} {\bibfnamefont {P.~R.}\ \bibnamefont
  {Zangara}}, \bibinfo {author} {\bibfnamefont {A.}~\bibnamefont {Wood}},
  \bibinfo {author} {\bibfnamefont {M.~W.}\ \bibnamefont {Doherty}},\ and\
  \bibinfo {author} {\bibfnamefont {C.~A.}\ \bibnamefont {Meriles}},\ }\href
  {https://doi.org/10.1103/PhysRevB.100.235410} {\bibfield  {journal} {\bibinfo
   {journal} {Phys. Rev. B}\ }\textbf {\bibinfo {volume} {100}},\ \bibinfo
  {pages} {235410} (\bibinfo {year} {2019})}\BibitemShut {NoStop}%
\bibitem [{\citenamefont {Henshaw}\ \emph {et~al.}(2019)\citenamefont
  {Henshaw}, \citenamefont {Pagliero}, \citenamefont {Zangara}, \citenamefont
  {Franzoni}, \citenamefont {Ajoy}, \citenamefont {Acosta}, \citenamefont
  {Reimer}, \citenamefont {Pines},\ and\ \citenamefont
  {Meriles}}]{henshaw_carbon-13_2019}%
  \BibitemOpen
  \bibfield  {author} {\bibinfo {author} {\bibfnamefont {J.}~\bibnamefont
  {Henshaw}}, \bibinfo {author} {\bibfnamefont {D.}~\bibnamefont {Pagliero}},
  \bibinfo {author} {\bibfnamefont {P.~R.}\ \bibnamefont {Zangara}}, \bibinfo
  {author} {\bibfnamefont {M.~B.}\ \bibnamefont {Franzoni}}, \bibinfo {author}
  {\bibfnamefont {A.}~\bibnamefont {Ajoy}}, \bibinfo {author} {\bibfnamefont
  {R.~H.}\ \bibnamefont {Acosta}}, \bibinfo {author} {\bibfnamefont {J.~A.}\
  \bibnamefont {Reimer}}, \bibinfo {author} {\bibfnamefont {A.}~\bibnamefont
  {Pines}},\ and\ \bibinfo {author} {\bibfnamefont {C.~A.}\ \bibnamefont
  {Meriles}},\ }\href {https://doi.org/10.1073/pnas.1908780116} {\bibfield
  {journal} {\bibinfo  {journal} {PNAS}\ }\textbf {\bibinfo {volume} {116}},\
  \bibinfo {pages} {18334} (\bibinfo {year} {2019})}\BibitemShut {NoStop}%
\bibitem [{\citenamefont {Childress}\ \emph {et~al.}(2006)\citenamefont
  {Childress}, \citenamefont {Gurudev~Dutt}, \citenamefont {Taylor},
  \citenamefont {Zibrov}, \citenamefont {Jelezko}, \citenamefont {Wrachtrup},
  \citenamefont {Hemmer},\ and\ \citenamefont
  {Lukin}}]{childress_coherent_2006}%
  \BibitemOpen
  \bibfield  {author} {\bibinfo {author} {\bibfnamefont {L.}~\bibnamefont
  {Childress}}, \bibinfo {author} {\bibfnamefont {M.~V.}\ \bibnamefont
  {Gurudev~Dutt}}, \bibinfo {author} {\bibfnamefont {J.~M.}\ \bibnamefont
  {Taylor}}, \bibinfo {author} {\bibfnamefont {A.~S.}\ \bibnamefont {Zibrov}},
  \bibinfo {author} {\bibfnamefont {F.}~\bibnamefont {Jelezko}}, \bibinfo
  {author} {\bibfnamefont {J.}~\bibnamefont {Wrachtrup}}, \bibinfo {author}
  {\bibfnamefont {P.~R.}\ \bibnamefont {Hemmer}},\ and\ \bibinfo {author}
  {\bibfnamefont {M.~D.}\ \bibnamefont {Lukin}},\ }\href
  {https://doi.org/10.1126/science.1131871} {\bibfield  {journal} {\bibinfo
  {journal} {Science}\ }\textbf {\bibinfo {volume} {314}},\ \bibinfo {pages}
  {281} (\bibinfo {year} {2006})}\BibitemShut {NoStop}%
\bibitem [{\citenamefont {Zhang}\ \emph {et~al.}(2023)\citenamefont {Zhang},
  \citenamefont {Joos}, \citenamefont {Bluvstein}, \citenamefont {Lyu},\ and\
  \citenamefont {Bleszynski~Jayich}}]{zhang_reporter-spin-assisted_2023}%
  \BibitemOpen
  \bibfield  {author} {\bibinfo {author} {\bibfnamefont {Z.}~\bibnamefont
  {Zhang}}, \bibinfo {author} {\bibfnamefont {M.}~\bibnamefont {Joos}},
  \bibinfo {author} {\bibfnamefont {D.}~\bibnamefont {Bluvstein}}, \bibinfo
  {author} {\bibfnamefont {Y.}~\bibnamefont {Lyu}},\ and\ \bibinfo {author}
  {\bibfnamefont {A.~C.}\ \bibnamefont {Bleszynski~Jayich}},\ }\href
  {https://doi.org/10.1103/PhysRevApplied.19.L031004} {\bibfield  {journal}
  {\bibinfo  {journal} {Phys. Rev. Appl.}\ }\textbf {\bibinfo {volume} {19}},\
  \bibinfo {pages} {L031004} (\bibinfo {year} {2023})}\BibitemShut {NoStop}%
\bibitem [{\citenamefont {Davis}\ \emph {et~al.}(2023)\citenamefont {Davis},
  \citenamefont {Ye}, \citenamefont {Machado}, \citenamefont {Meynell},
  \citenamefont {Wu}, \citenamefont {Mittiga}, \citenamefont {Schenken},
  \citenamefont {Joos}, \citenamefont {Kobrin}, \citenamefont {Lyu},
  \citenamefont {Wang}, \citenamefont {Bluvstein}, \citenamefont {Choi},
  \citenamefont {Zu}, \citenamefont {Jayich},\ and\ \citenamefont
  {Yao}}]{davis_probing_2023}%
  \BibitemOpen
  \bibfield  {author} {\bibinfo {author} {\bibfnamefont {E.~J.}\ \bibnamefont
  {Davis}}, \bibinfo {author} {\bibfnamefont {B.}~\bibnamefont {Ye}}, \bibinfo
  {author} {\bibfnamefont {F.}~\bibnamefont {Machado}}, \bibinfo {author}
  {\bibfnamefont {S.~A.}\ \bibnamefont {Meynell}}, \bibinfo {author}
  {\bibfnamefont {W.}~\bibnamefont {Wu}}, \bibinfo {author} {\bibfnamefont
  {T.}~\bibnamefont {Mittiga}}, \bibinfo {author} {\bibfnamefont
  {W.}~\bibnamefont {Schenken}}, \bibinfo {author} {\bibfnamefont
  {M.}~\bibnamefont {Joos}}, \bibinfo {author} {\bibfnamefont {B.}~\bibnamefont
  {Kobrin}}, \bibinfo {author} {\bibfnamefont {Y.}~\bibnamefont {Lyu}},
  \bibinfo {author} {\bibfnamefont {Z.}~\bibnamefont {Wang}}, \bibinfo {author}
  {\bibfnamefont {D.}~\bibnamefont {Bluvstein}}, \bibinfo {author}
  {\bibfnamefont {S.}~\bibnamefont {Choi}}, \bibinfo {author} {\bibfnamefont
  {C.}~\bibnamefont {Zu}}, \bibinfo {author} {\bibfnamefont {A.~C.~B.}\
  \bibnamefont {Jayich}},\ and\ \bibinfo {author} {\bibfnamefont {N.~Y.}\
  \bibnamefont {Yao}},\ }\href {https://doi.org/10.1038/s41567-023-01944-5}
  {\bibfield  {journal} {\bibinfo  {journal} {Nat. Phys.}\ }\textbf {\bibinfo
  {volume} {19}},\ \bibinfo {pages} {836} (\bibinfo {year} {2023})}\BibitemShut
  {NoStop}%
\bibitem [{\citenamefont {Park}\ \emph {et~al.}(2022)\citenamefont {Park},
  \citenamefont {Lee}, \citenamefont {Han}, \citenamefont {Oh},\ and\
  \citenamefont {Seo}}]{park_decoherence_2022}%
  \BibitemOpen
  \bibfield  {author} {\bibinfo {author} {\bibfnamefont {H.}~\bibnamefont
  {Park}}, \bibinfo {author} {\bibfnamefont {J.}~\bibnamefont {Lee}}, \bibinfo
  {author} {\bibfnamefont {S.}~\bibnamefont {Han}}, \bibinfo {author}
  {\bibfnamefont {S.}~\bibnamefont {Oh}},\ and\ \bibinfo {author}
  {\bibfnamefont {H.}~\bibnamefont {Seo}},\ }\href
  {https://doi.org/10.1038/s41534-022-00605-4} {\bibfield  {journal} {\bibinfo
  {journal} {npj Quantum Inf.}\ }\textbf {\bibinfo {volume} {8}},\ \bibinfo
  {pages} {1} (\bibinfo {year} {2022})}\BibitemShut {NoStop}%
\bibitem [{\citenamefont {Onizhuk}\ and\ \citenamefont
  {Galli}(2023)}]{onizhuk_bath-limited_2023}%
  \BibitemOpen
  \bibfield  {author} {\bibinfo {author} {\bibfnamefont {M.}~\bibnamefont
  {Onizhuk}}\ and\ \bibinfo {author} {\bibfnamefont {G.}~\bibnamefont
  {Galli}},\ }\href {https://doi.org/10.1103/PhysRevB.108.075306} {\bibfield
  {journal} {\bibinfo  {journal} {Phys. Rev. B}\ }\textbf {\bibinfo {volume}
  {108}},\ \bibinfo {pages} {075306} (\bibinfo {year} {2023})}\BibitemShut
  {NoStop}%
\bibitem [{\citenamefont {Ashfold}\ \emph {et~al.}(2020)\citenamefont
  {Ashfold}, \citenamefont {Goss}, \citenamefont {Green}, \citenamefont {May},
  \citenamefont {Newton},\ and\ \citenamefont
  {Peaker}}]{ashfold_nitrogen_2020}%
  \BibitemOpen
  \bibfield  {author} {\bibinfo {author} {\bibfnamefont {M.~N.~R.}\
  \bibnamefont {Ashfold}}, \bibinfo {author} {\bibfnamefont {J.~P.}\
  \bibnamefont {Goss}}, \bibinfo {author} {\bibfnamefont {B.~L.}\ \bibnamefont
  {Green}}, \bibinfo {author} {\bibfnamefont {P.~W.}\ \bibnamefont {May}},
  \bibinfo {author} {\bibfnamefont {M.~E.}\ \bibnamefont {Newton}},\ and\
  \bibinfo {author} {\bibfnamefont {C.~V.}\ \bibnamefont {Peaker}},\ }\href
  {https://doi.org/10.1021/acs.chemrev.9b00518} {\bibfield  {journal} {\bibinfo
   {journal} {Chem. Rev.}\ }\textbf {\bibinfo {volume} {120}},\ \bibinfo
  {pages} {5745} (\bibinfo {year} {2020})}\BibitemShut {NoStop}%
\bibitem [{\citenamefont {Deák}\ \emph {et~al.}(2014)\citenamefont {Deák},
  \citenamefont {Aradi}, \citenamefont {Kaviani}, \citenamefont {Frauenheim},\
  and\ \citenamefont {Gali}}]{deak_formation_2014}%
  \BibitemOpen
  \bibfield  {author} {\bibinfo {author} {\bibfnamefont {P.}~\bibnamefont
  {Deák}}, \bibinfo {author} {\bibfnamefont {B.}~\bibnamefont {Aradi}},
  \bibinfo {author} {\bibfnamefont {M.}~\bibnamefont {Kaviani}}, \bibinfo
  {author} {\bibfnamefont {T.}~\bibnamefont {Frauenheim}},\ and\ \bibinfo
  {author} {\bibfnamefont {A.}~\bibnamefont {Gali}},\ }\href
  {https://doi.org/10.1103/PhysRevB.89.075203} {\bibfield  {journal} {\bibinfo
  {journal} {Phys. Rev. B.}\ }\textbf {\bibinfo {volume} {89}},\ \bibinfo
  {pages} {075203} (\bibinfo {year} {2014})}\BibitemShut {NoStop}%
\bibitem [{\citenamefont {Ulbricht}\ \emph {et~al.}(2011)\citenamefont
  {Ulbricht}, \citenamefont {van~der Post}, \citenamefont {Goss}, \citenamefont
  {Briddon}, \citenamefont {Jones}, \citenamefont {Khan},\ and\ \citenamefont
  {Bonn}}]{ulbricht_single_2011}%
  \BibitemOpen
  \bibfield  {author} {\bibinfo {author} {\bibfnamefont {R.}~\bibnamefont
  {Ulbricht}}, \bibinfo {author} {\bibfnamefont {S.~T.}\ \bibnamefont {van~der
  Post}}, \bibinfo {author} {\bibfnamefont {J.~P.}\ \bibnamefont {Goss}},
  \bibinfo {author} {\bibfnamefont {P.~R.}\ \bibnamefont {Briddon}}, \bibinfo
  {author} {\bibfnamefont {R.}~\bibnamefont {Jones}}, \bibinfo {author}
  {\bibfnamefont {R.~U.~A.}\ \bibnamefont {Khan}},\ and\ \bibinfo {author}
  {\bibfnamefont {M.}~\bibnamefont {Bonn}},\ }\href
  {https://doi.org/10.1103/PhysRevB.84.165202} {\bibfield  {journal} {\bibinfo
  {journal} {Phys. Rev. B.}\ }\textbf {\bibinfo {volume} {84}},\ \bibinfo
  {pages} {165202} (\bibinfo {year} {2011})}\BibitemShut {NoStop}%
\bibitem [{\citenamefont {Davies}(1979)}]{davies_dynamic_1979}%
  \BibitemOpen
  \bibfield  {author} {\bibinfo {author} {\bibfnamefont {G.}~\bibnamefont
  {Davies}},\ }\href {https://doi.org/10.1088/0022-3719/12/13/019} {\bibfield
  {journal} {\bibinfo  {journal} {J. Phys. C: Solid State Phys.}\ }\textbf
  {\bibinfo {volume} {12}},\ \bibinfo {pages} {2551} (\bibinfo {year}
  {1979})}\BibitemShut {NoStop}%
\bibitem [{\citenamefont {Davies}(1981)}]{davies_jahn-teller_1981}%
  \BibitemOpen
  \bibfield  {author} {\bibinfo {author} {\bibfnamefont {G.}~\bibnamefont
  {Davies}},\ }\href {https://doi.org/10.1088/0034-4885/44/7/003} {\bibfield
  {journal} {\bibinfo  {journal} {Rep. Prog. Phys.}\ }\textbf {\bibinfo
  {volume} {44}},\ \bibinfo {pages} {787} (\bibinfo {year} {1981})}\BibitemShut
  {NoStop}%
\bibitem [{\citenamefont {Ammerlaan}\ and\ \citenamefont
  {Burgemeister}(1981)}]{ammerlaan_reorientation_1981}%
  \BibitemOpen
  \bibfield  {author} {\bibinfo {author} {\bibfnamefont {C.~A.~J.}\
  \bibnamefont {Ammerlaan}}\ and\ \bibinfo {author} {\bibfnamefont {E.~A.}\
  \bibnamefont {Burgemeister}},\ }\href
  {https://doi.org/10.1103/PhysRevLett.47.954} {\bibfield  {journal} {\bibinfo
  {journal} {Phys. Rev. Lett.}\ }\textbf {\bibinfo {volume} {47}},\ \bibinfo
  {pages} {954} (\bibinfo {year} {1981})}\BibitemShut {NoStop}%
\bibitem [{\citenamefont {Doherty}\ \emph {et~al.}(2016)\citenamefont
  {Doherty}, \citenamefont {Meriles}, \citenamefont {Alkauskas}, \citenamefont
  {Fedder}, \citenamefont {Sellars},\ and\ \citenamefont
  {Manson}}]{doherty_towards_2016}%
  \BibitemOpen
  \bibfield  {author} {\bibinfo {author} {\bibfnamefont {M.}~\bibnamefont
  {Doherty}}, \bibinfo {author} {\bibfnamefont {C.}~\bibnamefont {Meriles}},
  \bibinfo {author} {\bibfnamefont {A.}~\bibnamefont {Alkauskas}}, \bibinfo
  {author} {\bibfnamefont {H.}~\bibnamefont {Fedder}}, \bibinfo {author}
  {\bibfnamefont {M.}~\bibnamefont {Sellars}},\ and\ \bibinfo {author}
  {\bibfnamefont {N.}~\bibnamefont {Manson}},\ }\href
  {https://doi.org/10.1103/PhysRevX.6.041035} {\bibfield  {journal} {\bibinfo
  {journal} {Phys. Rev. X.}\ }\textbf {\bibinfo {volume} {6}},\ \bibinfo
  {pages} {041035} (\bibinfo {year} {2016})}\BibitemShut {NoStop}%
\bibitem [{\citenamefont {Bauch}\ \emph {et~al.}(2020)\citenamefont {Bauch},
  \citenamefont {Singh}, \citenamefont {Lee}, \citenamefont {Hart},
  \citenamefont {Schloss}, \citenamefont {Turner}, \citenamefont {Barry},
  \citenamefont {Pham}, \citenamefont {Bar-Gill}, \citenamefont {Yelin},\ and\
  \citenamefont {Walsworth}}]{bauch_decoherence_2020}%
  \BibitemOpen
  \bibfield  {author} {\bibinfo {author} {\bibfnamefont {E.}~\bibnamefont
  {Bauch}}, \bibinfo {author} {\bibfnamefont {S.}~\bibnamefont {Singh}},
  \bibinfo {author} {\bibfnamefont {J.}~\bibnamefont {Lee}}, \bibinfo {author}
  {\bibfnamefont {C.~A.}\ \bibnamefont {Hart}}, \bibinfo {author}
  {\bibfnamefont {J.~M.}\ \bibnamefont {Schloss}}, \bibinfo {author}
  {\bibfnamefont {M.~J.}\ \bibnamefont {Turner}}, \bibinfo {author}
  {\bibfnamefont {J.~F.}\ \bibnamefont {Barry}}, \bibinfo {author}
  {\bibfnamefont {L.~M.}\ \bibnamefont {Pham}}, \bibinfo {author}
  {\bibfnamefont {N.}~\bibnamefont {Bar-Gill}}, \bibinfo {author}
  {\bibfnamefont {S.~F.}\ \bibnamefont {Yelin}},\ and\ \bibinfo {author}
  {\bibfnamefont {R.~L.}\ \bibnamefont {Walsworth}},\ }\href
  {https://doi.org/10.1103/PhysRevB.102.134210} {\bibfield  {journal} {\bibinfo
   {journal} {Phys. Rev. B.}\ }\textbf {\bibinfo {volume} {102}},\ \bibinfo
  {pages} {134210} (\bibinfo {year} {2020})}\BibitemShut {NoStop}%
\bibitem [{\citenamefont {Maze}\ \emph {et~al.}(2008)\citenamefont {Maze},
  \citenamefont {Taylor},\ and\ \citenamefont {Lukin}}]{maze_electron_2008}%
  \BibitemOpen
  \bibfield  {author} {\bibinfo {author} {\bibfnamefont {J.~R.}\ \bibnamefont
  {Maze}}, \bibinfo {author} {\bibfnamefont {J.~M.}\ \bibnamefont {Taylor}},\
  and\ \bibinfo {author} {\bibfnamefont {M.~D.}\ \bibnamefont {Lukin}},\ }\href
  {https://doi.org/10.1103/PhysRevB.78.094303} {\bibfield  {journal} {\bibinfo
  {journal} {Phys. Rev. B}\ }\textbf {\bibinfo {volume} {78}},\ \bibinfo
  {pages} {094303} (\bibinfo {year} {2008})}\BibitemShut {NoStop}%
\bibitem [{\citenamefont {Stanwix}\ \emph {et~al.}(2010)\citenamefont
  {Stanwix}, \citenamefont {Pham}, \citenamefont {Maze}, \citenamefont
  {Le~Sage}, \citenamefont {Yeung}, \citenamefont {Cappellaro}, \citenamefont
  {Hemmer}, \citenamefont {Yacoby}, \citenamefont {Lukin},\ and\ \citenamefont
  {Walsworth}}]{stanwix_coherence_2010}%
  \BibitemOpen
  \bibfield  {author} {\bibinfo {author} {\bibfnamefont {P.~L.}\ \bibnamefont
  {Stanwix}}, \bibinfo {author} {\bibfnamefont {L.~M.}\ \bibnamefont {Pham}},
  \bibinfo {author} {\bibfnamefont {J.~R.}\ \bibnamefont {Maze}}, \bibinfo
  {author} {\bibfnamefont {D.}~\bibnamefont {Le~Sage}}, \bibinfo {author}
  {\bibfnamefont {T.~K.}\ \bibnamefont {Yeung}}, \bibinfo {author}
  {\bibfnamefont {P.}~\bibnamefont {Cappellaro}}, \bibinfo {author}
  {\bibfnamefont {P.~R.}\ \bibnamefont {Hemmer}}, \bibinfo {author}
  {\bibfnamefont {A.}~\bibnamefont {Yacoby}}, \bibinfo {author} {\bibfnamefont
  {M.~D.}\ \bibnamefont {Lukin}},\ and\ \bibinfo {author} {\bibfnamefont
  {R.~L.}\ \bibnamefont {Walsworth}},\ }\href
  {https://doi.org/10.1103/PhysRevB.82.201201} {\bibfield  {journal} {\bibinfo
  {journal} {Phys. Rev. B}\ }\textbf {\bibinfo {volume} {82}},\ \bibinfo
  {pages} {201201} (\bibinfo {year} {2010})}\BibitemShut {NoStop}%
\bibitem [{\citenamefont {Viola}\ \emph {et~al.}(1999)\citenamefont {Viola},
  \citenamefont {Knill},\ and\ \citenamefont {Lloyd}}]{viola_dynamical_1999}%
  \BibitemOpen
  \bibfield  {author} {\bibinfo {author} {\bibfnamefont {L.}~\bibnamefont
  {Viola}}, \bibinfo {author} {\bibfnamefont {E.}~\bibnamefont {Knill}},\ and\
  \bibinfo {author} {\bibfnamefont {S.}~\bibnamefont {Lloyd}},\ }\href
  {https://doi.org/10.1103/PhysRevLett.82.2417} {\bibfield  {journal} {\bibinfo
   {journal} {Phys. Rev. Lett.}\ }\textbf {\bibinfo {volume} {82}},\ \bibinfo
  {pages} {2417} (\bibinfo {year} {1999})}\BibitemShut {NoStop}%
\bibitem [{\citenamefont {Ryan}\ \emph {et~al.}(2010)\citenamefont {Ryan},
  \citenamefont {Hodges},\ and\ \citenamefont {Cory}}]{ryan_robust_2010}%
  \BibitemOpen
  \bibfield  {author} {\bibinfo {author} {\bibfnamefont {C.~A.}\ \bibnamefont
  {Ryan}}, \bibinfo {author} {\bibfnamefont {J.~S.}\ \bibnamefont {Hodges}},\
  and\ \bibinfo {author} {\bibfnamefont {D.~G.}\ \bibnamefont {Cory}},\ }\href
  {https://doi.org/10.1103/PhysRevLett.105.200402} {\bibfield  {journal}
  {\bibinfo  {journal} {Phys. Rev. Lett.}\ }\textbf {\bibinfo {volume} {105}},\
  \bibinfo {pages} {200402} (\bibinfo {year} {2010})}\BibitemShut {NoStop}%
\bibitem [{\citenamefont {Naydenov}\ \emph {et~al.}(2011)\citenamefont
  {Naydenov}, \citenamefont {Dolde}, \citenamefont {Hall}, \citenamefont
  {Shin}, \citenamefont {Fedder}, \citenamefont {Hollenberg}, \citenamefont
  {Jelezko},\ and\ \citenamefont {Wrachtrup}}]{naydenov_dynamical_2011}%
  \BibitemOpen
  \bibfield  {author} {\bibinfo {author} {\bibfnamefont {B.}~\bibnamefont
  {Naydenov}}, \bibinfo {author} {\bibfnamefont {F.}~\bibnamefont {Dolde}},
  \bibinfo {author} {\bibfnamefont {L.~T.}\ \bibnamefont {Hall}}, \bibinfo
  {author} {\bibfnamefont {C.}~\bibnamefont {Shin}}, \bibinfo {author}
  {\bibfnamefont {H.}~\bibnamefont {Fedder}}, \bibinfo {author} {\bibfnamefont
  {L.~C.~L.}\ \bibnamefont {Hollenberg}}, \bibinfo {author} {\bibfnamefont
  {F.}~\bibnamefont {Jelezko}},\ and\ \bibinfo {author} {\bibfnamefont
  {J.}~\bibnamefont {Wrachtrup}},\ }\href
  {https://doi.org/10.1103/PhysRevB.83.081201} {\bibfield  {journal} {\bibinfo
  {journal} {Phys. Rev. B.}\ }\textbf {\bibinfo {volume} {83}},\ \bibinfo
  {pages} {081201} (\bibinfo {year} {2011})}\BibitemShut {NoStop}%
\bibitem [{\citenamefont {Gullion}\ \emph {et~al.}(1990)\citenamefont
  {Gullion}, \citenamefont {Baker},\ and\ \citenamefont
  {Conradi}}]{gullion_new_1990}%
  \BibitemOpen
  \bibfield  {author} {\bibinfo {author} {\bibfnamefont {T.}~\bibnamefont
  {Gullion}}, \bibinfo {author} {\bibfnamefont {D.~B.}\ \bibnamefont {Baker}},\
  and\ \bibinfo {author} {\bibfnamefont {M.~S.}\ \bibnamefont {Conradi}},\
  }\href {https://doi.org/10.1016/0022-2364(90)90331-3} {\bibfield  {journal}
  {\bibinfo  {journal} {J. Magn. Reson.}\ }\textbf {\bibinfo {volume} {89}},\
  \bibinfo {pages} {479} (\bibinfo {year} {1990})}\BibitemShut {NoStop}%
\bibitem [{\citenamefont {Carr}\ and\ \citenamefont
  {Purcell}(1954)}]{carr_effects_1954}%
  \BibitemOpen
  \bibfield  {author} {\bibinfo {author} {\bibfnamefont {H.~Y.}\ \bibnamefont
  {Carr}}\ and\ \bibinfo {author} {\bibfnamefont {E.~M.}\ \bibnamefont
  {Purcell}},\ }\href {https://doi.org/10.1103/PhysRev.94.630} {\bibfield
  {journal} {\bibinfo  {journal} {Phys. Rev.}\ }\textbf {\bibinfo {volume}
  {94}},\ \bibinfo {pages} {630} (\bibinfo {year} {1954})}\BibitemShut
  {NoStop}%
\bibitem [{\citenamefont {Meiboom}\ and\ \citenamefont
  {Gill}(1958)}]{meiboom_modified_1958}%
  \BibitemOpen
  \bibfield  {author} {\bibinfo {author} {\bibfnamefont {S.}~\bibnamefont
  {Meiboom}}\ and\ \bibinfo {author} {\bibfnamefont {D.}~\bibnamefont {Gill}},\
  }\href {https://doi.org/10.1063/1.1716296} {\bibfield  {journal} {\bibinfo
  {journal} {Rev. Sci. Instrum.}\ }\textbf {\bibinfo {volume} {29}},\ \bibinfo
  {pages} {688} (\bibinfo {year} {1958})}\BibitemShut {NoStop}%
\bibitem [{\citenamefont {Hall}\ \emph {et~al.}(2014)\citenamefont {Hall},
  \citenamefont {Cole},\ and\ \citenamefont {Hollenberg}}]{hall_analytic_2014}%
  \BibitemOpen
  \bibfield  {author} {\bibinfo {author} {\bibfnamefont {L.~T.}\ \bibnamefont
  {Hall}}, \bibinfo {author} {\bibfnamefont {J.~H.}\ \bibnamefont {Cole}},\
  and\ \bibinfo {author} {\bibfnamefont {L.~C.~L.}\ \bibnamefont
  {Hollenberg}},\ }\href {https://doi.org/10.1103/PhysRevB.90.075201}
  {\bibfield  {journal} {\bibinfo  {journal} {Phys. Rev. B.}\ }\textbf
  {\bibinfo {volume} {90}},\ \bibinfo {pages} {075201} (\bibinfo {year}
  {2014})}\BibitemShut {NoStop}%
\bibitem [{\citenamefont {Hall}\ \emph {et~al.}(2016)\citenamefont {Hall},
  \citenamefont {Kehayias}, \citenamefont {Simpson}, \citenamefont {Jarmola},
  \citenamefont {Stacey}, \citenamefont {Budker},\ and\ \citenamefont
  {Hollenberg}}]{hall_detection_2016}%
  \BibitemOpen
  \bibfield  {author} {\bibinfo {author} {\bibfnamefont {L.~T.}\ \bibnamefont
  {Hall}}, \bibinfo {author} {\bibfnamefont {P.}~\bibnamefont {Kehayias}},
  \bibinfo {author} {\bibfnamefont {D.~A.}\ \bibnamefont {Simpson}}, \bibinfo
  {author} {\bibfnamefont {A.}~\bibnamefont {Jarmola}}, \bibinfo {author}
  {\bibfnamefont {A.}~\bibnamefont {Stacey}}, \bibinfo {author} {\bibfnamefont
  {D.}~\bibnamefont {Budker}},\ and\ \bibinfo {author} {\bibfnamefont
  {L.~C.~L.}\ \bibnamefont {Hollenberg}},\ }\href
  {https://doi.org/10.1038/ncomms10211} {\bibfield  {journal} {\bibinfo
  {journal} {Nat. Commun.}\ }\textbf {\bibinfo {volume} {7}},\ \bibinfo {pages}
  {10211} (\bibinfo {year} {2016})}\BibitemShut {NoStop}%
\bibitem [{\citenamefont {Wood}\ \emph {et~al.}(2021)\citenamefont {Wood},
  \citenamefont {Goldblatt}, \citenamefont {Anderson}, \citenamefont
  {Hollenberg}, \citenamefont {Scholten},\ and\ \citenamefont
  {Martin}}]{wood_anisotropic_2021}%
  \BibitemOpen
  \bibfield  {author} {\bibinfo {author} {\bibfnamefont {A.~A.}\ \bibnamefont
  {Wood}}, \bibinfo {author} {\bibfnamefont {R.~M.}\ \bibnamefont {Goldblatt}},
  \bibinfo {author} {\bibfnamefont {R.~P.}\ \bibnamefont {Anderson}}, \bibinfo
  {author} {\bibfnamefont {L.~C.~L.}\ \bibnamefont {Hollenberg}}, \bibinfo
  {author} {\bibfnamefont {R.~E.}\ \bibnamefont {Scholten}},\ and\ \bibinfo
  {author} {\bibfnamefont {A.~M.}\ \bibnamefont {Martin}},\ }\href
  {https://doi.org/10.1103/PhysRevB.104.085419} {\bibfield  {journal} {\bibinfo
   {journal} {Phys. Rev. B.}\ }\textbf {\bibinfo {volume} {104}},\ \bibinfo
  {pages} {085419} (\bibinfo {year} {2021})}\BibitemShut {NoStop}%
\bibitem [{\citenamefont {Laraoui}\ \emph {et~al.}(2013)\citenamefont
  {Laraoui}, \citenamefont {Dolde}, \citenamefont {Burk}, \citenamefont
  {Reinhard}, \citenamefont {Wrachtrup},\ and\ \citenamefont
  {Meriles}}]{laraoui_high-resolution_2013}%
  \BibitemOpen
  \bibfield  {author} {\bibinfo {author} {\bibfnamefont {A.}~\bibnamefont
  {Laraoui}}, \bibinfo {author} {\bibfnamefont {F.}~\bibnamefont {Dolde}},
  \bibinfo {author} {\bibfnamefont {C.}~\bibnamefont {Burk}}, \bibinfo {author}
  {\bibfnamefont {F.}~\bibnamefont {Reinhard}}, \bibinfo {author}
  {\bibfnamefont {J.}~\bibnamefont {Wrachtrup}},\ and\ \bibinfo {author}
  {\bibfnamefont {C.~A.}\ \bibnamefont {Meriles}},\ }\href
  {https://www.nature.com/articles/ncomms2685} {\bibfield  {journal} {\bibinfo
  {journal} {Nat. Commun.}\ }\textbf {\bibinfo {volume} {4}} (\bibinfo {year}
  {2013})}\BibitemShut {NoStop}%
\bibitem [{\citenamefont {Hubrich}\ \emph {et~al.}(1997)\citenamefont
  {Hubrich}, \citenamefont {Bauer},\ and\ \citenamefont
  {Spiess}}]{hubrich_magic-angle_1997}%
  \BibitemOpen
  \bibfield  {author} {\bibinfo {author} {\bibfnamefont {M.}~\bibnamefont
  {Hubrich}}, \bibinfo {author} {\bibfnamefont {C.}~\bibnamefont {Bauer}},\
  and\ \bibinfo {author} {\bibfnamefont {H.~W.}\ \bibnamefont {Spiess}},\
  }\href {https://doi.org/10.1016/S0009-2614(97)00562-9} {\bibfield  {journal}
  {\bibinfo  {journal} {Chem. Phys. Lett.}\ }\textbf {\bibinfo {volume}
  {273}},\ \bibinfo {pages} {259} (\bibinfo {year} {1997})}\BibitemShut
  {NoStop}%
\bibitem [{\citenamefont {Wood}\ \emph {et~al.}(2017)\citenamefont {Wood},
  \citenamefont {Lilette}, \citenamefont {Fein}, \citenamefont {Perunicic},
  \citenamefont {Hollenberg}, \citenamefont {Scholten},\ and\ \citenamefont
  {Martin}}]{wood_magnetic_2017}%
  \BibitemOpen
  \bibfield  {author} {\bibinfo {author} {\bibfnamefont {A.~A.}\ \bibnamefont
  {Wood}}, \bibinfo {author} {\bibfnamefont {E.}~\bibnamefont {Lilette}},
  \bibinfo {author} {\bibfnamefont {Y.~Y.}\ \bibnamefont {Fein}}, \bibinfo
  {author} {\bibfnamefont {V.~S.}\ \bibnamefont {Perunicic}}, \bibinfo {author}
  {\bibfnamefont {L.~C.~L.}\ \bibnamefont {Hollenberg}}, \bibinfo {author}
  {\bibfnamefont {R.~E.}\ \bibnamefont {Scholten}},\ and\ \bibinfo {author}
  {\bibfnamefont {A.~M.}\ \bibnamefont {Martin}},\ }\href
  {https://doi.org/10.1038/nphys4221} {\bibfield  {journal} {\bibinfo
  {journal} {Nat. Phys.}\ }\textbf {\bibinfo {volume} {13}},\ \bibinfo {pages}
  {1070} (\bibinfo {year} {2017})}\BibitemShut {NoStop}%
\bibitem [{\citenamefont {Jin}\ \emph {et~al.}(2023)\citenamefont {Jin},
  \citenamefont {Shen}, \citenamefont {Ju}, \citenamefont {Gao}, \citenamefont
  {Zu}, \citenamefont {Grine},\ and\ \citenamefont {Li}}]{jin_quantum_2023}%
  \BibitemOpen
  \bibfield  {author} {\bibinfo {author} {\bibfnamefont {Y.}~\bibnamefont
  {Jin}}, \bibinfo {author} {\bibfnamefont {K.}~\bibnamefont {Shen}}, \bibinfo
  {author} {\bibfnamefont {P.}~\bibnamefont {Ju}}, \bibinfo {author}
  {\bibfnamefont {X.}~\bibnamefont {Gao}}, \bibinfo {author} {\bibfnamefont
  {C.}~\bibnamefont {Zu}}, \bibinfo {author} {\bibfnamefont {A.~J.}\
  \bibnamefont {Grine}},\ and\ \bibinfo {author} {\bibfnamefont
  {T.}~\bibnamefont {Li}},\ }\Eprint {https://arxiv.org/abs/2309.05821}
  {arXiv:2309.05821 [quant-ph]}  (\bibinfo {year} {2023})\BibitemShut {NoStop}%
\bibitem [{\citenamefont {Ungar}\ \emph {et~al.}(2023)\citenamefont {Ungar},
  \citenamefont {Cappellaro}, \citenamefont {Cooper},\ and\ \citenamefont
  {Sun}}]{ungar_identification_2023}%
  \BibitemOpen
  \bibfield  {author} {\bibinfo {author} {\bibfnamefont {A.}~\bibnamefont
  {Ungar}}, \bibinfo {author} {\bibfnamefont {P.}~\bibnamefont {Cappellaro}},
  \bibinfo {author} {\bibfnamefont {A.}~\bibnamefont {Cooper}},\ and\ \bibinfo
  {author} {\bibfnamefont {W.~K.~C.}\ \bibnamefont {Sun}},\ }\Eprint
  {https://arxiv.org/abs/2306.17155} {arXiv:2306.17155 [quant-ph]}  (\bibinfo
  {year} {2023})\BibitemShut {NoStop}%
\bibitem [{\citenamefont {Grotz}\ \emph {et~al.}(2011)\citenamefont {Grotz},
  \citenamefont {Beck}, \citenamefont {Neumann}, \citenamefont {Naydenov},
  \citenamefont {Reuter}, \citenamefont {Reinhard}, \citenamefont {Jelezko},
  \citenamefont {Wrachtrup}, \citenamefont {Schweinfurth}, \citenamefont
  {Sarkar},\ and\ \citenamefont {Hemmer}}]{grotz_sensing_2011}%
  \BibitemOpen
  \bibfield  {author} {\bibinfo {author} {\bibfnamefont {B.}~\bibnamefont
  {Grotz}}, \bibinfo {author} {\bibfnamefont {J.}~\bibnamefont {Beck}},
  \bibinfo {author} {\bibfnamefont {P.}~\bibnamefont {Neumann}}, \bibinfo
  {author} {\bibfnamefont {B.}~\bibnamefont {Naydenov}}, \bibinfo {author}
  {\bibfnamefont {R.}~\bibnamefont {Reuter}}, \bibinfo {author} {\bibfnamefont
  {F.}~\bibnamefont {Reinhard}}, \bibinfo {author} {\bibfnamefont
  {F.}~\bibnamefont {Jelezko}}, \bibinfo {author} {\bibfnamefont
  {J.}~\bibnamefont {Wrachtrup}}, \bibinfo {author} {\bibfnamefont
  {D.}~\bibnamefont {Schweinfurth}}, \bibinfo {author} {\bibfnamefont
  {B.}~\bibnamefont {Sarkar}},\ and\ \bibinfo {author} {\bibfnamefont
  {P.}~\bibnamefont {Hemmer}},\ }\href
  {https://doi.org/10.1088/1367-2630/13/5/055004} {\bibfield  {journal}
  {\bibinfo  {journal} {New J. Phys.}\ }\textbf {\bibinfo {volume} {13}},\
  \bibinfo {pages} {055004} (\bibinfo {year} {2011})}\BibitemShut {NoStop}%
\bibitem [{\citenamefont {Salikhov}\ \emph {et~al.}(1981)\citenamefont
  {Salikhov}, \citenamefont {Dzuba},\ and\ \citenamefont
  {Raitsimring}}]{salikhov_theory_1981}%
  \BibitemOpen
  \bibfield  {author} {\bibinfo {author} {\bibfnamefont {K.~M.}\ \bibnamefont
  {Salikhov}}, \bibinfo {author} {\bibfnamefont {S.~A.}\ \bibnamefont
  {Dzuba}},\ and\ \bibinfo {author} {\bibfnamefont {A.~M.}\ \bibnamefont
  {Raitsimring}},\ }\href {https://doi.org/10.1016/0022-2364(81)90216-X}
  {\bibfield  {journal} {\bibinfo  {journal} {J. Magn. Reson.}\ }\textbf
  {\bibinfo {volume} {42}},\ \bibinfo {pages} {255} (\bibinfo {year}
  {1981})}\BibitemShut {NoStop}%
\bibitem [{\citenamefont {Stepanov}\ and\ \citenamefont
  {Takahashi}(2016)}]{stepanov_determination_2016}%
  \BibitemOpen
  \bibfield  {author} {\bibinfo {author} {\bibfnamefont {V.}~\bibnamefont
  {Stepanov}}\ and\ \bibinfo {author} {\bibfnamefont {S.}~\bibnamefont
  {Takahashi}},\ }\href {https://doi.org/10.1103/PhysRevB.94.024421} {\bibfield
   {journal} {\bibinfo  {journal} {Phys. Rev. B.}\ }\textbf {\bibinfo {volume}
  {94}},\ \bibinfo {pages} {024421} (\bibinfo {year} {2016})}\BibitemShut
  {NoStop}%
\bibitem [{\citenamefont {Edmonds}\ \emph {et~al.}(2012)\citenamefont
  {Edmonds}, \citenamefont {D’Haenens-Johansson}, \citenamefont {Cruddace},
  \citenamefont {Newton}, \citenamefont {Fu}, \citenamefont {Santori},
  \citenamefont {Beausoleil}, \citenamefont {Twitchen},\ and\ \citenamefont
  {Markham}}]{edmonds_production_2012}%
  \BibitemOpen
  \bibfield  {author} {\bibinfo {author} {\bibfnamefont {A.~M.}\ \bibnamefont
  {Edmonds}}, \bibinfo {author} {\bibfnamefont {U.~F.~S.}\ \bibnamefont
  {D’Haenens-Johansson}}, \bibinfo {author} {\bibfnamefont {R.~J.}\
  \bibnamefont {Cruddace}}, \bibinfo {author} {\bibfnamefont {M.~E.}\
  \bibnamefont {Newton}}, \bibinfo {author} {\bibfnamefont {K.-M.~C.}\
  \bibnamefont {Fu}}, \bibinfo {author} {\bibfnamefont {C.}~\bibnamefont
  {Santori}}, \bibinfo {author} {\bibfnamefont {R.~G.}\ \bibnamefont
  {Beausoleil}}, \bibinfo {author} {\bibfnamefont {D.~J.}\ \bibnamefont
  {Twitchen}},\ and\ \bibinfo {author} {\bibfnamefont {M.~L.}\ \bibnamefont
  {Markham}},\ }\href {https://doi.org/10.1103/PhysRevB.86.035201} {\bibfield
  {journal} {\bibinfo  {journal} {Phys. Rev. B.}\ }\textbf {\bibinfo {volume}
  {86}},\ \bibinfo {pages} {035201} (\bibinfo {year} {2012})}\BibitemShut
  {NoStop}%
\end{thebibliography}

\end{document}